\documentclass[a4paper,11pt]{article}
\usepackage[utf8]{inputenc}
\usepackage{jheppub}
\usepackage[parfill]{parskip}    
\usepackage{graphicx}
\usepackage{dcolumn}
\usepackage{amsmath,amssymb,amsfonts,braket}
\usepackage{bm}
\usepackage{epstopdf}
\usepackage{hyperref}
\usepackage{mathtools}
\usepackage{bigfoot}
\usepackage{setspace}
\usepackage{soul,xcolor}
\usepackage{mathrsfs}

\newcommand{\tfabc} {{\tilde f}^{{\tilde a}{\tilde b}{\tilde c}}}
\newcommand{\fabc}{f^{abc}} 
\newcommand{\ta}{\tilde{a}}

\newcommand{\al}{\alpha}

\newcommand{\ft}{\tilde{f}}

\newcommand{\be}{\begin{equation}}
\newcommand{\ee}{\end{equation}}
\newcommand{\bea}{\begin{eqnarray}}
\newcommand{\eea}{\end{eqnarray}}
\newcommand{\bsp}{\begin{split}}
\newcommand{\ens}{\end{split}}

\newcommand{\fofr}{f(r)}

\newcommand{\half}{\textstyle\frac{1}{2}}

\newcommand{\del}{\partial}

\title{\boldmath The classical double copy in curved spacetimes: Perturbative Yang-Mills from the bi-adjoint scalar }
\author{Siddharth G. Prabhu}
\affiliation{International Centre for Theoretical Sciences, Tata Institute of Fundamental Research, Shivakote, Bengaluru 560089, India.}
\emailAdd{siddharth.g.prabhu@gmail.com}
\abstract{We formulate a version of the double copy for classical fields in curved spacetimes. We provide a correspondence between perturbative solutions to the bi-adjoint scalar equations and those of the Yang-Mills equations in position space. At the linear level, we show that there exists a map between these solutions for maximally symmetric spacetime backgrounds, that provides every Yang-Mills solution by the action of an appropriate differential operator on a bi-adjoint scalar solution. Given the existence of a linearized map, we show that it is possible to cast the solutions of the Yang-Mills equations at arbitrary perturbation order in terms of the corresponding bi-adjoint scalar solutions. This all-order map is reminiscent of the flat space BCJ double copy, and works for any curved spacetime where the perturbative expansion holds. We show that these results have the right flat space limit, and that the correspondence is agnostic to the choice of gauge.}

\begin{document}
\maketitle

\section{Introduction}
\label{sec:intro}
The problem of computing solutions to classical field equations around a background spacetime is a very interesting and considerably challenging endeavour. The complexity increases with the spin of the field, and the degree of non-linearity of the equations. In recent years, a promising approach towards simplifying such an analysis has come about from an unexpected arena, namely the study of scattering amplitudes in string theory and quantum field theory. 

Kawai, Lewellen and Tye (KLT) showed that a duality between open and closed strings provides a relation between perturbative Yang-Mills and gravity amplitudes in the field theory limit. Tree-level gravity scattering amplitudes can be obtained from their much simpler Yang-Mills counterparts \cite{Kawai:1985xq}. A more direct and general construction with far-reaching consequences was provided purely within quantum field theory by the work of Bern, Carrasco and Johansson~\cite{Bern:2008qj,Bern:2010ue}. Using certain replacements of colour factors with kinematics ones, they obtained tree-level gravity scattering amplitudes from two copies of Yang-Mills scattering amplitudes, hence dubbing this procedure the \emph{double copy}~\cite{Bern:2010yg}. There are various proofs of the double copy at the tree level~\cite{BjerrumBohr:2010hn,Mafra:2011kj,Bjerrum-Bohr:2016axv}, where it is seen to be equivalent to the KLT relations. There is also extensive and ever-increasing evidence at the loop level leading up to ~\cite{Bern:2017ucb,Bern:2018jmv}.
Moreover, an increasingly large set of theories have been shown to be related by such a procedure. A comprehensive review of the duality between colour and kinematics, the double-copy relations, as well as references to the growing body of literature on this subject can be found in~\cite{Carrasco:2015iwa,Borsten:2020bgv,Bern:2019prr}. 

These results point to relationships between dissimilar theories, and also provide a route to simplifying computations in more complicated theories by casting them as computations in simpler ones. This line of thought has turned out to be fruitful even at the classical level. It allows the recasting of solutions of classical field equations in terms of those with lesser spin, and lesser degree of non-linearity. Various approaches to the classical double copy have appeared in the literature. One idea is to relate exact solutions in Yang-Mills and in gravity using Kerr-Schild coordinates~\cite{Monteiro:2014cda}, which has been generalized to many different settings in~\cite{Luna:2015paa,Luna:2016due,Luna:2016hge,Luna:2017dtq,Bahjat-Abbas:2017htu,Carrillo-Gonzalez:2017iyj,Berman:2018hwd,Lee:2018gxc,Gurses:2018ckx,CarrilloGonzalez:2019gof,Andrzejewski:2019hub,Cho:2019ype,Alawadhi:2019urr,Kim:2019jwm,Bah:2019sda,Arkani-Hamed:2019ymq,Lescano:2020nve,Elor:2020nqe,Luna:2020adi,Easson:2020esh,Moynihan:2020ejh,Alfonsi:2020lub,Gumus:2020hbb,Keeler:2020rcv,Bahjat-Abbas:2020cyb}. One could, however, imagine generating perturbative solutions of classical gravity without actually solving Einstein's equations.  Such an approach, formulated in  ~\cite{Goldberger:2016iau} and developed in \cite{Goldberger:2017frp,Goldberger:2017vcg,Goldberger:2017ogt,Chester:2017vcz,Li:2018qap,Shen:2018ebu, Carrillo-Gonzalez:2018pjk, Carrillo-Gonzalez:2019aao, Goldberger:2019xef}, arrives at classical gravitational radiation from a set of sources travelling along their worldlines.\footnote{See Fig. (1) in~\cite{Carrillo-Gonzalez:2018pjk} for a sketch of different theories related by the BCJ double copy, and also by the exact and perturbative avatars of the classical one.}

An entirely different approach is the study of quantum amplitudes to extract classical observables and to derive the classical double copy \cite{Kosower:2018adc,Maybee:2019jus,delaCruz:2020bbn,Johansson:2019dnu,Johansson:2014zca,Plefka:2019wyg,Bautista:2019tdr,Bautista:2019evw,PV:2019uuv}. Some other approaches to the classical double copy are the use of convolution \cite{Cardoso:2016ngt,Cardoso:2016amd,Borsten:2020xbt,Anastasiou:2014qba,Anastasiou:2018rdx}, spinorial approaches  \cite{Luna:2018dpt, Alawadhi:2020jrv}, the self-dual double copy \cite{Monteiro:2011pc,Chacon:2020fmr}, and the perturbiner expansion \cite{Mizera:2018jbh}. A very useful application of the classical double copy is in the simplification of calculations for binary inspirals relevant for gravitational wave detections \cite{Cheung:2018wkq,Bern:2019nnu,Bern:2019crd}, see also \cite{Almeida:2020mrg, Kalin:2019rwq,Kalin:2020mvi,Kalin:2020fhe}. Another interesting avenue is its relation with electromagnetic duality \cite{Huang:2019cja, Banerjee:2019saj,Alawadhi:2019urr} .

Most investigations of the double copy have been conducted in flat space. While the Kerr-Schild approach was generalized to curved spacetimes in~\cite{Luna:2015paa,Bahjat-Abbas:2017htu,Carrillo-Gonzalez:2017iyj,Gumus:2020hbb}, (see also \cite{Casali:2020vuy}), one might wonder if the double copy can help with gravitational perturbation theory around curved backgrounds. This was studied for plane wave background spacetimes in~\cite{Adamo:2017nia,Ilderton:2018lsf,Adamo:2018mpq,Adamo:2019zmk,Adamo:2020qru}.
When thinking about the double copy in curved spaces, one could ask whether the background spacetime is itself double copied. Indeed, some curved spacetimes in the Kerr-Schild gauge have been shown to be double copies of gauge theories. So, one can imagine getting both the background and the perturbation from a double copy procedure. Instead, we wish to ask whether we can relate perturbative solutions about the same background spacetime. 

We formulate a new avatar of the classical double copy for curved spacetimes. The first rung of the double copy is occupied by the theory of the bi-adjoint scalar whose usefulness in the classical context has been explored in \cite{Goldberger:2017frp}, and also in \cite{White:2016jzc,DeSmet:2017rve,Bahjat-Abbas:2018vgo}. Here, we present a correspondence between pure bi-adjoint scalar solutions and those of pure Yang-Mills in position space. Around maximally symmetric spacetime backgrounds, we find that every linearized Yang-Mills solution can be obtained from its bi-adjoint counterpart. Furthermore, we find that any such linearized map can be lifted to all orders in perturbation, using ideas from the flat space BCJ double copy. The all-order map is independent of the map at the linearized level, and works for any background spacetime where the perturbative expansion is valid. The correspondence has the novel feature that it works in any gauge and for all solutions of the classical equations. We will treat the Yang-Mills to gravity correspondence elsewhere, where we will show that a similar map generates every linearized solution of the Einstein's equations in a maximally symmetric spacetime from a linearized solution of the Yang-Mills equations.

\paragraph{\bf Outline of the method and summary \\}

We will now outline the schematics of our procedure that generates a perturbative solution to the Yang-Mills equations beginning from any perturbative solution of the bi-adjoint scalar about a fixed background spacetime, and provide all the details of this construction in the subsequent sections. 

We consider solutions of the fields about a $(d+1)$ dimensional background spacetime, that admits a $M_{2} \times S^{d-1}$ decomposition,
\begin{equation} \label{eq:Bgdecomp}
ds^2=g_{\mu \nu} \, dx^{\mu} dx^{\nu}= g_{\alpha \beta} \, dy^{\alpha} dy^{\beta} +r^2(y)  d\Omega_{d-1}^2,
\end{equation}

where $d\Omega_{d-1}^2=\gamma_{ij}d\theta^i d\theta^j$, with $\gamma_{ij}$ being the standard round metric on $S^{d-1}$. Here, $(\mu,\nu)$ run over the indices of the full $(d+1)$ dimensional spacetime; $(\alpha,\beta)$ run over the two directions of $M_2$; and $(i,j)$ run over the $(d-1)$ directions of $S^{d-1}$. The covariant derivatives associated with these spaces shall be denoted by $\nabla_\mu$, $\hat{\nabla}_{\alpha}$, and $D_i$ respectively. 

We employ the spherical symmetry to decompose any function $F(y_\alpha, \theta_i)$ on this spacetime into its scalar, vector and tensor components with respect to rotations of $S^{d-1}$.  Schematically,
\bea \label{eq:shdecomp}
F(y_\alpha, \theta_i)=\sum_{k} f_{k} (y_\alpha) \, H_{k}(\theta_i), 
\eea
where the sum is over appropriate $S^{d-1}$ spherical harmonics $ H_{k}(\theta_i)$ with eigenvalue $k$ (see Appendix (\ref{sec:spharms}) for more details), and $f_k$ are scalar functions with respect to $S^{d-1}$.
In particular, solutions of the bi-adjoint scalar, Yang-Mills and Einstein equations can all be decomposed into their scalar, vector and tensor components in this manner, using spherical harmonics of rank up to two. The advantage of using such a decomposition lies in the fact that the scalar, vector and tensor harmonics span inequivalent representations of the rotation group. So, any rotationally invariant operator does not mix the different components of the spherical harmonic decomposition. As a result, the field equations can be independently studied in each of these sectors - scalar, vector and tensor. Such decompositions have been used to arrive at a gauge invariant formalism for gravitational perturbations in \cite{Kodama:2000fa,Kodama:2003jz}.

We begin with studying the bi-adjoint scalar field on this background spacetime. The bi-adjoint scalar $\Phi_{a \tilde{a}}$ transforms in the adjoint representation of two independent global groups, $G$ and $\tilde{G}$, and satisfies the equations of motion
\begin{equation} \label{eq:baeoms}
\nabla^{\mu}\nabla_\mu \Phi_{a \tilde{a}}= - \fabc \tfabc \, \Phi_{b \tilde{b}}\, \Phi_{c \tilde{c}}
\end{equation}
We will study these equations in a perturbative expansion for the bi-adjoint field in the coupling constant $y$ 
\begin{equation}
\Phi_{a \tilde{a}}= \sum_{n} y^n \phi_{a \ta}^{(n)},
\end{equation}
where $ \phi_{a \ta}^{(i)}$ admits a decomposition of the kind in (Eq. \ref{eq:shdecomp}) 
\begin{equation} \label{eq:badecomp}
\phi_{a \ta}^{(n)}(y_\alpha,\theta_i)=\sum_k c_{a \ta}^{(n)}\, \phi_{k}^{(n)}(y_\alpha) S_{k}(\theta_i).
\end{equation}
Here, the $c_{a \ta}^{(n)}$ are color factors arising from $G\times \tilde{G}$, and the $S_{k}(\theta_i)$ are scalar spherical harmonics on $S^{(d-1)}$. For convenience, we drop the $k$ labels and explicit coordinate dependences, and keep the sum over $k$ implicit, so that (Eq. \ref{eq:badecomp}) can be written simply as 
\begin{equation} \label{eq:badecomps}
\phi_{a \ta}^{(n)}= c_{a \ta}^{(n)}\, \phi^{(n)} \, S, 
\end{equation}
At each perturbative order $n$, the equations of motion (Eq.\ref{eq:baeoms}) are reduced to solving a differential equation on $M_2$ for $\phi^{(n)}$. For maximally symmetric spacetimes, we write down the equation at the linear order and solve it in Sec. \ref{sec:balin}.

Given these solutions $\phi_{a \ta}^{(n)}$, we wish to generate solutions to the Yang-Mills equations of motion
\begin{equation}
\nabla^{\mu} F^{a}_{\mu \nu}= - \fabc \, A_{b}^{\mu}\, F^{c}_{\mu \nu}
\end{equation}
in a perturbative expansion in the Yang-Mills coupling constant $g$,
\begin{equation}
A^{a}_{\mu}= \sum_{n} g^n \mathcal{A}^{a(n)}_{\mu}.
\end{equation}

We shall see that the map between the bi-adjoint solutions and Yang-Mills solutions can be written in any gauge. In order to write down explicit solutions for the gauge field, we will find it convenient to choose a gauge. The Yang-Mills field can be decomposed as in (Eq. \ref{eq:shdecomp}), with both scalar and vector harmonics making an appearance. Only the scalar harmonics $S$ appear in the scalar components of the gauge field $\mathcal{A}_{a,\alpha}$. We choose the gauge in which only the vector harmonics $V_i$ appear in the vector components of the gauge field $\mathcal{A}_{a,i}$. We will show that, in this gauge, the Yang-Mills field can be written as
\begin{equation}
\begin{split}
\mathcal{A}_{a, \mu}^{(n)}
&=  \phi^{V(n)}_{a}\, V_i \, \delta_{\mu}^{i} +  \frac{1}{r^{d-3}} \epsilon_{\alpha \beta}\left(\hat{\nabla}^{\beta}\phi^{S(n)}_{a}\right)  \, S \, \delta_{\mu}^{\alpha}.  
\end{split}
\end{equation}

At every perturbative order, the Yang-Mills equations can then be rewritten as two-dimensional equations on $M_2$ for the two scalar functions $\phi^{S(n)}_{a}$ and $\phi^{V(n)}_{a}$. In Sec. \ref{sec:ymlin}, we provide all the details of this perturbative setup for solving the Yang-Mills equations. We also write down the equations at the linear order for maximally symmetric spacetimes and solve them. The corresponding result for the Maxwell field in AdS was obtained in~\cite{Ishibashi:2004wx}.

The desired map between the solutions for the bi-adjoint equations of motion and those of the Yang-Mills equation is achieved by providing operators $\mathcal{K}_{a,\mu}^{(n)}$ \, such that 
\begin{equation} \label{eq:ordernmap}
\mathcal{A}_{a, \mu}^{(n)}=  \mathcal{K}_{a,\mu}^{(n)} \left[\phi^{(n)}\right]
\end{equation}
In the formalism we have described, this amounts to generating $\phi^{S(n)}_{a}$ and $\phi^{V(n)}_{a}$, given the solutions to the bi-adjoint equations of motion $\phi^{(n)}$. At the first order, the color dependence is trivial. In Sec.~\ref{sec:linmap}, we will find differential operators that generate the linearized Yang-Mills solutions about maximally symmetric spacetimes from the linearized bi-adjoint solutions. We also show that this map can be written down in any gauge by phrasing it in terms of the gauge invariants at the linear level. 

Given any linearized map, and not necessarily the one we describe, we write down a formal map that generates perturbative Yang-Mills solutions at arbitrary order from the corresponding bi-adjoint scalar solutions in Sec. \ref{sec:allordermap}. Assuming the validity of the perturbative theory, the all-order map works for any background spacetime, and is independent of the linearized map, or the gauge choice. In Sec. \ref{sec:allordermink}, we specialize the all-order map to Minkowski spacetimes where it simplifies, and provides explicit expressions for the gauge field. In Sec. \ref{sec:allorderAdS}, we describe the procedure to implement this map in $AdS$ spacetimes.

\section{The linearized bi-adjoint to Yang-Mills correspondence}
\label{sec:linmapintro}
In this section, we shall first write down the solutions to the linearized bi-adjoint scalar equations and the linearized Yang-Mills equations. The procedure we outline works for the general background spacetime given in Eq. (\ref{eq:Bgdecomp}). In order to provide explicit solutions however, we specialize the $M_{2} \times S^{d-1}$ decomposition of the background spacetime to the coordinates
\begin{equation}\begin{split} \label{eq:adsglobal}
ds^2 =-f(r)\, dt^2+\frac{dr^2}{f(r)}  + r^2 d\Omega^2_{d-1} \ 
\end{split}\end{equation} 
Further, we shall write solutions to these equations only for global AdS where $f(r)=(1+\frac{r^2}{R^2})$, and global dS where $f(r)=(1- \frac{r^2}{R^2})$. In appendix \ref{sec:geneq}, we show how the flat space limit of these solutions can be recovered. 

We then provide the linearized map Eq. (\ref{eq:ordernmap}) at the first order. In other words, we find simple differential operators whose action on a linearized solution to the bi-adjoint scalar equations of motion in $AdS$ and $dS$ generates a solution to the corresponding linearized Yang-Mills equations. We see that any solution of linearized Yang-Mills equations can be obtained in this manner.

\subsection{The bi-adjoint scalar theory and its linearized solutions}
\label{sec:balin}

We shall consider the theory of a massless scalar $\Phi_{a \tilde{a}}$ that transforms bi-linearly in the adjoint representation of two independent global groups $G \times \tilde{G}$, and has cubic self interactions only.

The action for this theory is 
\begin{equation}
\label{eq:bis}
{\cal S}_\Phi=\int d^{d+1}x \, \sqrt{g} \left(\frac{1}{2}(\nabla^\mu \Phi^{a {\ta}})(\nabla_\mu \Phi_{a {\tilde a}}) + \frac{1}{3} f^{abc} \ft^{{\tilde a} {\tilde b} {\tilde c}} \Phi_{a {\tilde a}} \Phi_{b {\tilde b}} \Phi_{c {\tilde c}} \right), 
\end{equation}
varying which gives the equations of motion (Eq. \ref{eq:baeom}).
\begin{equation} \label{eq:baeom}
\nabla^{\mu}\nabla_\mu \Phi_{a \tilde{a}}= - \fabc \tfabc \, \Phi_{b \tilde{b}}\, \Phi_{c \tilde{c}}
\end{equation}

We use the perturbative ansatz 
\begin{equation} \label{eq:bapert}
\Phi_{a \tilde{a}}= \sum_{n} y^n \phi_{a \ta}^{(n)}.
\end{equation}
At the linearized level, the decomposition Eq. (\ref{eq:badecomp}) we employ becomes
\begin{equation} \label{eq:badecomplin}
\phi_{a \ta}^{(1)}(t,r,\theta_i)= t_a \, t_{\ta} \, \sum_{\ell}   \phi_{\ell}^{(1)}(t,r)\, S_{\ell}(\theta_i),
\end{equation}
where $t_a,t_{\ta}$ are the generators of the gauge groups $G,\tilde{G}$ respectively.
 
Using the defining equation of the scalar spherical harmonics given in Appendix (\ref{sec:spharms}), $\phi_{\ell}^{(1)}$ satisfies
\begin{equation} \label{eq:bahyp}
\left( \frac{1}{r^{d-1}} \del_r \left(r^{d-1}\,\fofr\, \del_r \right) -\frac{1}{\fofr} \frac{\partial^2}{\partial t^2}- \frac{l(l+d-2)}{r^2} \right) \phi_{\ell}^{(1)}=0
\end{equation}
In global AdS or dS, this is a hypergeometric equation whose solutions can be written in terms of the Gaussian hypergeometric functions.

We shall write down the solutions in frequency space $\phi_{\Delta,l}(r)$ via $\phi_{\ell}^{(1)}(t,r)=\sum_{\Delta} \, e^{-i \frac{\Delta}{R} t}\phi_{\Delta,l}(r)$.
The solution that is generically regular at the origin can be written as
\begin{equation} \label{eq:baregsoln}
r^{l} \, \fofr^{\frac{\Delta}{2}}  \, _2F_1\left(\half \left(l+\Delta\right),\half \left(d+l+\Delta\right) \,;\frac{d}{2}+l;-\frac{r^2}{R^2}\right),
\end{equation}
whereas the solution that generically diverges at the origin is
\begin{equation} \label{eq:bairregsoln} 
\frac{1}{r^{d+l-2}}\, \fofr^{\frac{\Delta}{2}} \,_2F_1\left(\half \left(2-l+\Delta\right),\half \left(2-d-l+\Delta\right) \,;2-\frac{d}{2}-l;-\frac{r^2}{R^2}\right).  \\ 
\end{equation}
We will simply use $\phi_{\Delta,l}$ to denote either of these solutions. We will be reducing the Yang-Mills equations to equations for a set of scalar variables. We shall use appropriate superscripts for them to differentiate them from the ones in this section. When the background is flat, Eq. (\ref{eq:bahyp}) reduces to a Bessel equation. In Appendix \ref{sec:geneq}, we show how the appropriate flat limit of the solutions in this section reduce to Bessel functions.

\subsection{Yang-Mills theory and its linearized solutions}
\label{sec:ymlin}

The Yang-Mills equations of motion in a curved background can be written as
\begin{equation}
\nabla^{\mu} F^{a}_{\mu \nu}= - \fabc \, A_{b}^{\mu}\, F^{c}_{\mu \nu}
\end{equation}

We want to look at the solutions of these equations in a perturbative expansion in the Yang-Mills coupling constant $g$ 
\begin{equation}
A^{a}_{\mu}= \sum_{n} g^n \mathcal{A}^{a(n)}_{\mu}.
\end{equation}
We adapt the spherical harmonic decomposition Eq. (\ref{eq:shdecomp}) for the Yang-Mills field at each perturbative order. In order to do so, we examine how various components of the field behave under the $S^{d-1}$ rotations. The components $\mathcal{A}^{a(n)}_{\alpha}$ along the two dimensional $M_2$ directions  transform as scalars under the $S^{d-1}$ rotations, which means that only scalar spherical harmonics will show up in their decomposition. The components of the field  $\mathcal{A}^{a(n)}_{i}$ which are along the $S^{d-1}$ directions will have contributions both from the scalar as well as the vector spherical harmonics. So, the spherical harmonic decomposition for the Yang-Mills field can be written as follows
\begin{equation}\label{eq:GaugeDecomp}
\begin{split}
\mathcal{A}^{a(n)}_{\alpha}(y_\alpha,\theta_i)&= \sum_{\ell} f^{a(n)}_{\alpha,\ell}(y_\alpha)\, S_{\ell}(\theta_i) \\
\mathcal{A}^{a(n)}_{i}(y_\alpha,\theta_i)&= \sum_{\ell} \left(f^{a(n)}_{S,\ell}(y_\alpha)\, D_{i} S_{\ell}(\theta_i) + f^{a(n)}_{V,\ell}(y_\alpha) V_{\ell,i}(\theta_i) \right),
\end{split}
\end{equation}
where the subscripts $S$ and $V$ denote scalar and vector respectively. Under a gauge transformation $\chi^{a(n)}$, the field transforms as $ \mathcal{A}^{a(n)}_{\mu} \rightarrow \mathcal{A}^{a(n)}_{\mu} + \nabla_\mu \chi^{a(n)}$. The gauge function can itself be decomposed in spherical harmonics as 
\begin{equation}\label{eq:gaugetrans}
\chi^{a(n)}(y_\alpha,\theta_i)= \sum_{\ell} h^{a(n)}_{\ell}(y_\alpha) \, S_{\ell}(\theta_i)
\end{equation}
We shall use this gauge freedom to only have vector spherical harmonics in the decomposition of $\mathcal{A}^{a(n)}_{i}$, i.e. to put all the $f^{a(n)}_{S,\ell}=0$. We also simplify our notation by dropping the coordinate dependences, the $\ell$ labels, and suppressing the sum over all allowed values of $\ell$.  In the gauge we have employed, the spherical harmonic decomposition of the Yang-Mills field in the global coordinates  of Eq.~(\ref{eq:adsglobal}) can be written as
\begin{equation} \label{eq:ymgauge_allorder}
\begin{split}
\mathcal{A}^{a(n)}_{t}&=  \fofr \, \frac{S}{r^{d-3}} \frac{\partial}{\partial r}\phi^{S(n)}_{a}, \\
\mathcal{A}^{a(n)}_{r}&=  \frac{1}{\fofr} \, \frac{S}{r^{d-3}} \frac{\partial}{\partial t}\phi^{S(n)}_{a}, \\
\mathcal{A}^{a(n)}_{i}&=  V_i \, \phi^{V(n)}_{a},
\end{split}
\end{equation}
where we have introduced the two functions $\phi^{S(n)}_{a}$ and $\phi^{V(n)}_{a}$, with superscripts which denote that they appear along with the scalar and vector harmonics respectively. This notation differentiates them from the bi-adjoint scalar variables, which are written without any such superscripts. 
For future convenience, we shall rewrite this as advertised in the introduction
\begin{equation}\label{eq:GaugetoSc}
\begin{split}
\mathcal{A}_{a, \mu}^{(n)}
&=  \phi^{V(n)}_{a}\, V_i \, \delta_{\mu}^{i} +   \frac{1}{r^{d-3}} \epsilon_{\alpha \beta}\left(\hat{\nabla}^{\beta}\phi^{S(n)}_{a}\right)  \, S \, \delta_{\mu}^{\alpha}, 
\end{split}
\end{equation}
where $\epsilon_{\alpha \beta}$ is the two-dimensional Levi-Civita symbol. 

At the linearized level, the equations of motion are $\nabla_\nu F^{\mu \nu}_{a}=0$. The color dependence is trivial. Writing $\phi_a^{S(1)}=t_a \phi^S$ and $\phi_a^{V(1)}=t_a \phi^V$, we can write the linearized equations of motion as
\begin{equation}\begin{split} \label{maxwelleqns}
\nabla_\nu F_a^{t\nu} &= t_a \frac{S}{r^{d-1}}\frac{\partial}{\partial r} r^2\left\{r^{d-3}\frac{\partial }{\partial r}\left( \frac{\fofr}{r^{d-3}} \frac{\partial \phi^S}{\partial r}\right)-\frac{\ell(\ell+d-2)}{r^2}\phi^S-
\frac{1}{\fofr}  \frac{\partial^2 \phi^S}{\partial t^2} \right\}\ ,\\
\nabla_\nu F_a^{\nu r} &= t_a \frac{S}{r^{d-1}}\frac{\partial}{\partial t} r^2\left\{r^{d-3}\frac{\partial }{\partial r}\left( \frac{\fofr}{r^{d-3}} \frac{\partial \phi^S}{\partial r}\right)-\frac{\ell(\ell+d-2)}{r^2}\phi^S-
\frac{1}{\fofr} \frac{\partial^2 \phi^S}{\partial t^2} \right\}\ ,\\
\nabla_\nu F_a^{i\nu} &= -t_a \frac{V^i}{r^2} \left\{\frac{1}{r^{d-3}}\frac{\partial }{\partial r}\left( \fofr \, r^{d-3} \frac{\partial \phi^V}{\partial r}\right)-\frac{(\ell+1)(\ell+d-3)}{r^2}\phi^V-
\frac{1}{\fofr}\frac{\partial^2 \phi^V}{\partial t^2} \right\}. \end{split}\end{equation}

Thus, we conclude that linearized Yang-Mills equations are satisfied provided \footnote{It might seem that we could have added $\frac{\lambda}{r^2}$ to the right side of the equation for $\phi^S$ in Eq.(\ref{eq:ymsc}), where $\lambda$ is a constant independent of $(r,t)$. However, the definition of $\phi^S$ via Eq. \ref{eq:ymgauge_allorder} is invariant if $\phi^S \rightarrow \phi^S + c$, where $c$ is another constant independent of $r$ and $t$. We use this freedom to choose $\lambda$ to be $0$.}
\begin{equation}\begin{split} \label{eq:ymsc}
r^{d-3}\frac{\partial }{\partial r}\left( \frac{\fofr}{r^{d-3}} \frac{\partial \phi^S}{\partial r}\right)-\frac{\ell(\ell+d-2)}{r^2}\phi^S-
\frac{1}{\fofr}  \frac{\partial^2 \phi^S}{\partial t^2} &= 0\ ,\\
\frac{1}{r^{d-3}}\frac{\partial }{\partial r}\left( \fofr \, r^{d-3} \frac{\partial \phi^V}{\partial r}\right)-\frac{(\ell+1)(\ell+d-3)}{r^2}\phi^V-
\frac{1}{\fofr}\frac{\partial^2 \phi^V}{\partial t^2} &= 0\ . \end{split}\end{equation}

The explicit solutions can be written in terms of hypergeometric functions in the frequency domain via $\phi^S_{\ell}(t,r)=r^{d-4}\,\sum_{\Delta}e^{-i\frac{\Delta}{R} t}\, \phi^S_{\Delta,l}(r)$ and $\phi^V_{\ell}(t,r)=\sum_{\Delta}e^{-i\frac{\Delta}{R} t}\, \phi^V_{\Delta,l}(r)$. The solutions regular at the origin are given by 
\begin{equation}\begin{split} \label{eq:ymregsolns}
\phi^{S}_{\Delta,\ell}(r) &=  r^{\ell+2} \, \fofr^{\frac{\Delta}{2}} \, {}_2 F_1\left(\frac{1}{2}(\Delta+\ell+2),\frac{1}{2}(\Delta+\ell+d-2),\frac{d}{2}+\ell, -\frac{r^2}{R^2}\right)\ ,
\\
\phi^{V}_{\Delta,\ell}(r) &=  r^{\ell+1} \, \fofr^{\frac{\Delta}{2}} \, {}_2 F_1\left(\frac{1}{2}(\Delta+\ell+1),\frac{1}{2}(\Delta+\ell+d-1),\frac{d}{2}+\ell, -\frac{r^2}{R^2}\right)\ . 
\end{split}\end{equation}
The solutions regular at infinity are given by

\begin{equation}\begin{split} \label{eq:ymirregsolns}
\phi^{S}_{\Delta,\ell}(r)&=\frac{1}{r^{\ell+d-4}} \, \fofr^{\frac{\Delta}{2}} \, {}_2 F_1\left(\frac{1}{2}(\Delta-\ell),\frac{1}{2}(\Delta-\ell-d+4),2-\frac{d}{2}-\ell, -\frac{r^2}{R^2}\right)\ ,
\\
\phi^{V}_{\Delta,\ell}(r)&= \frac{1}{r^{\ell+d-3}} \, \fofr^{\frac{\Delta}{2}} \, {}_2 F_1\left(\frac{1}{2}(\Delta-\ell+1),\frac{1}{2}(\Delta-\ell-d+3),2-\frac{d}{2}-\ell, -\frac{r^2}{R^2}\right)\ . \end{split}\end{equation}

In the rest of this paper, we use $\phi^{S}_{\Delta,\ell}(r)$ and $\phi^{S}_{\Delta,\ell}(r)$ to denote either of the respective solutions above. 

The equations obtained in~(\ref{eq:ymsc}) are again hypergeometric differential equations of the kind obtained in Sec. \ref{sec:balin}. In \cite{Ishibashi:2004wx}, the Maxwell equations in AdS were written down in this formalism, and after a change of variables, agree with the ones above.  We shall treat the linearized Einstein equations elsewhere (see \cite{Ishibashi:2004wx} too), but we just note here that they reduce to such equations too. In Appendix \ref{sec:geneq}, we treat the generalized equation of this kind and write down its solutions. Taking the flat space of the solutions appropriately, we show how Bessel functions are recovered.

While we have employed a convenient gauge choice, we can write the linearized solutions in a general gauge as well. In order to do this, we shall write the gauge invariants at the linearized order. We write the decomposition of the gauge field in Eq. (\ref{eq:GaugeDecomp}) as
\begin{equation}
\begin{split}
\mathcal{A}^{a(1)}_{\alpha}&= f^{a(1)}_{\alpha}\, S \\
\mathcal{A}^{a(1)}_{i}&= f^{a(1)}_{S}\, D_{i} S + f^{a(1)}_{V} V_{i} ,
\end{split}
\end{equation}
where, as before, we have dropped all coordinate dependencies, and $\ell$ labels. Since the gauge field enjoys only a scalar function worth of gauge freedom, it follows that its purely vector part $f^{a(1)}_{V}$ is gauge invariant by itself. Looking at the action of the gauge transformation Eq. (\ref{eq:gaugetrans}), it is clear that the other two gauge invariants on $M_2$ are given by $(f^{a(1)}_{\al} - \nabla_{\al} \, f^{a(1)}_{S})$. Hence, on the full spacetime, we can form the following gauge invariant combinations
\begin{equation}
\begin{split}
\mathcal{B}^{a}_{\alpha}&= \left(f^{a(1)}_{\al} - \nabla_{\al} \, f^{a(1)}_{S}\right)\, S \\
\mathcal{B}^{a}_{i}&= f^{a(1)}_{V} \, V_{i}.
\end{split}
\end{equation}
Now, we can simply write the gauge invariant analogue of Eq. (\ref{eq:GaugetoSc}) as  
\begin{equation}\label{eq:GaugeInvarToSc}
\begin{split}
\mathcal{B}_{a, \mu}
&=  \phi^{V(1)}_{a}\, V_i \, \delta_{\mu}^{i} +   \frac{1}{r^{d-3}} \epsilon_{\alpha \beta}\left(\hat{\nabla}^{\beta}\phi^{S(1)}_{a}\right)  \, S \, \delta_{\mu}^{\alpha}, 
\end{split}
\end{equation}
where $\phi^{S(1)}_{a}$ and $\phi^{V(1)}_{a}$ are the same scalar variables as before satisfying Eqs. (\ref{eq:ymsc}). Thus, the replacement $\mathcal{A}_{a, \mu}^{(1)} \rightarrow \mathcal{B}_{a, \mu}$ takes the results we get in our gauge to a general gauge. We can explicitly write down the general linearized solution for the gauge field as
\begin{equation} \label{eq:ym_lineargeneralgauge}
\begin{split}
\mathcal{A}^{a(1)}_{t}&=  \left\{\frac{\fofr}{r^{d-3}} \frac{\partial}{\partial r}\phi^{S(1)}_{a} +\nabla_{t} \, f^{a(1)}_{S}\right\} S, \\
\mathcal{A}^{a(n)}_{r}&=  \left\{\frac{1}{\fofr} \, \frac{1}{r^{d-3}} \frac{\partial}{\partial t}\phi^{S(1)}_{a} +\nabla_{r} \, f^{a(1)}_{S}\right\}S, \\
\mathcal{A}^{a(n)}_{i}&=   \phi^{V(1)}_{a} \, V_i +f^{a(1)}_{S} D_i S,
\end{split}
\end{equation}
with $f^{a(1)}_{S}(t,r)$ an arbitrary function, the choice of which determines the gauge one is in. Of course, this general solution can also be obtained by performing a gauge transformation on the solution in  Eq. (\ref{eq:GaugetoSc}) with  $f^{a(1)}_{S}(t,r)=0$.

\subsection{The bi-adjoint to Yang-Mills correspondence at first order}
\label{sec:linmap}

We wish to show that there exists a map between linearized bi-adjoint solutions and linearized Yang-Mills solutions. In the previous section, we have expressed the solutions of the Yang-Mills equations in terms of two functions, one each in the scalar and vector sectors, given in Eqs. (\ref{eq:ymregsolns}, \ref{eq:ymirregsolns}). Here, we show that these functions can be arrived at by applying a differential operator to the linearized bi-adjoint solutions in Eqs. (\ref{eq:baregsoln}, \ref{eq:bairregsoln}). This then implies the desired map. 

As we have noted, all these solutions are hypergeometric functions, and so finding a relation between them is an exercise in the theory of these functions. A natural place to look are the contiguous relations between hypergeometric functions which enable one to raise or lower the three parameters $(a,b,c)$ by integer values. However, here we need to raise or lower these parameters by half-integer values. We solve this problem by using the following construction. We first define linear differential operators that raise or lower the angular momentum $l$ by one unit. Applying such an operator once to the linearized bi-adjoint solution $\phi$ generates the function appearing in the vector sector of the linearized Yang-Mills solution $\phi^V$. Applying it again generates the function appearing in the scalar sector $\phi^S$. Starting with a bi-adjoint solution, we can either apply the raising or lowering operators to generate the Yang-Mills scalars of larger or smaller angular momentum respectively. Conversely, every Yang-Mills solution can be generated in this manner, starting with some bi-adjoint scalar solution. Instead of angular momentum, we can also choose to raise or lower in the frequency $\Delta$. All such identities which relate the linearized bi-adjoint solutions to the Yang-Mills ones can be obtained as consequences of the contiguous relations satisfied by the corresponding hypergeometric functions.

First, we generate the Yang-Mills scalars by lowering angular momentum,

\begin{equation}\begin{split} \label{eq:lowerl}
\phi^V_{\Delta,\ell}&= \left(\left(\ell+d-1\right)+r\, \fofr\, \frac{\partial}{\partial r}\right) \phi_{\Delta,\ell+1} \ ,\\
\phi^S_{\Delta,\ell}&= \left(\left(\ell+d-2\right)+r\, \fofr\,\frac{\partial}{\partial r}\right)\phi^V_{\Delta,\ell +1}\\
&=\left(\left(\ell+d-2\right)+r\, \fofr\,\frac{\partial}{\partial r}\right)\left(\left(\ell+d\right)+r\, \fofr\,\frac{\partial}{\partial r}\right) \phi_{\Delta,\ell+2} 
\end{split}\end{equation} 

Defining a \emph{lowering} operator in angular momentum space, $a_{\ell+1}\equiv \left\{\left(\ell+d-1\right)+r\, \fofr\, \frac{\partial}{\partial r}\right\}$, we can write these relations succinctly as
\begin{equation}\begin{split} 
\phi^V_{\Delta,\ell}&=a_{\ell+1} \, \phi_{\Delta,\ell+1} \ ,\\
\phi^S_{\Delta,\ell}&=a_{\ell}\,\phi^V_{\Delta,\ell+1}\\
&=a_{\ell} \,a_{\ell+2} \, \phi_{\Delta,\ell+2}
\end{split}\end{equation}
We note that the angular momentum number of these raising operators is never smaller than $0$, so that this is a definition for $a_{k}$ with $k\geq 0$. Also, this procedure gives $(\phi^S_{\Delta,\ell},\phi^V_{\Delta,\ell})$ for every possible value of $(\Delta,\ell)$, which implies that this procedure can be used to obtain all the linearized Yang-Mills solutions. 

We now write the analogous relations if we raise the angular momentum instead,
\begin{equation}\begin{split} 
\phi^V_{\Delta,\ell}&= \left(\left(\ell-1\right)-r\, \fofr\, \frac{\partial}{\partial r}\right) \phi_{\Delta,\ell-1} \ ,\\
\phi^S_{\Delta,\ell}&= \left(\ell-r\, \fofr\,\frac{\partial}{\partial r}\right)\phi^{V}_{\Delta,\ell -1}\\
&=\left(\ell-r\, \fofr\,\frac{\partial}{\partial r}\right)\left(\left(\ell-2\right)-r\, \fofr\,\frac{\partial}{\partial r}\right) \phi_{\Delta,\ell-2} 
\end{split}\end{equation} 

In the same vein as earlier, we can define a \emph{raising} operator in angular momentum space as
$a_{-\ell-1}=\left\{\left(\ell-1\right)-r\, \fofr\, \frac{\partial}{\partial r}\right\}$, and cast these relations as
\begin{equation}\begin{split} 
\phi^V_{\Delta,\ell}&=a_{-\ell-1} \, \phi_{\Delta,\ell-1} \ ,\\
\phi^S_{\Delta,\ell}&=a_{-\ell-2}\,\phi^{V}_{\Delta,\ell-1}\\
&=a_{-\ell-2} \,a_{-\ell} \, \phi_{\Delta,\ell-2}.
\end{split}\end{equation}
We note that the raising operators so defined never have their angular momentum index be equal to or greater than $0$. So taken together with the definition of the lowering operators, we have a definition of $a_k$ for all integers. The negative integers provide the raising operators, and the non-negative integers provide the lowering operators. 

The results of this section taken together with the ones of the previous sections, imply that for every linearized solution of the bi-adjoint equations of motion, we can generate a solution to the linearized Yang-Mills equations. Conversely, every linearized Yang-Mills solution can be obtained starting from the linearized bi-adjoint scalar solutions. We shall now explicitly write this map, found by raising or lowering angular momentum. In doing so, we will go back to using the time coordinate $t$ instead of dealing with the frequency $\Delta$. Restoring the coordinate dependencies, every solution to the linearized bi-adjoint equations of motion
\begin{equation} \label{eq:balinsoln}
\phi_{a \ta}^{(1)}(t,r,\theta_i)= t_a \, t_{\ta} \, \sum_{\ell}   \phi_{\ell}^{(1)}(t,r) S_{\ell}(\theta_i)
\end{equation}
generates the following solution to the linearized Yang-Mills equations of motion
\begin{equation} \label{eq:linmapone}
\begin{split}
\mathcal{A}^{a(1)}_{t}(t,r,\theta_i) &= t_a \,  \frac{\fofr}{r^{d-3}}\sum_{\ell}    \frac{\partial  }{\partial r}\left(r^{d-4} \,a_{\ell} \,a_{\ell+2} \,  \phi_{\ell+2}^{(1)}(t,r)\right) S_{\ell}(\theta_i), \, \\
\mathcal{A}^{a(1)}_{r}(t,r,\theta_i)&= t_a \,  \frac{1}{r^{d-3}\,\fofr}\sum_{\ell}  \frac{\partial  }{\partial t}\left(r^{d-4}a_{\ell} \,a_{\ell+2} \,  \phi_{\ell+2}^{(1)}(t,r)\right) S_{\ell}(\theta_i), \\
\mathcal{A}^{a(1)}_{i}(t,r,\theta_i)&= t_a  \sum_{\ell}   \,  a_{\ell+1} \, \phi_{\ell+1}^{(1)}(t,r) V_{\ell,i}(\theta_i).\\
\end{split}
\end{equation}
Instead, using raising operators, we can generate the following solution to the linearized Yang-Mills equations
\begin{equation} \label{eq:linmaptwo}
\begin{split}
\mathcal{A}^{a(1)}_{t}(t,r,\theta_i) &= t_a \,  \frac{\fofr}{r^{d-3}}\sum_{\ell}    \frac{\partial  }{\partial r}\left(r^{d-4} \,a_{-\ell} \, a_{-\ell-2} \, \phi_{\ell-2}^{(1)}(t,r)\right) S_{\ell}(\theta_i), \, \\
\mathcal{A}^{a(1)}_{r}(t,r,\theta_i)&= t_a \,  \frac{1}{r^{d-3}\,\fofr}\sum_{\ell}  \frac{\partial  }{\partial t}\left(r^{d-4}\,a_{-\ell} \, a_{-\ell-2} \, \phi_{\ell-2}^{(1)}(t,r)\right) S_{\ell}(\theta_i), \\
\mathcal{A}^{a(1)}_{i}(t,r,\theta_i)&= t_a  \sum_{\ell}   \, a_{-\ell-1} \,  \phi_{\ell-1}^{(1)}(t,r) V_{\ell,i}(\theta_i).\\
\end{split}
\end{equation}
Either of Eqs.(\ref{eq:linmapone},\ref{eq:linmaptwo}) provide the explicit map at the linearized level that we sought. 

The notation that we have introduced for the raising and lowering operators enables us to write a compact formula for the linearized map. Dropping the coordinate dependencies for convenience, this takes the form
\begin{equation} \label{eq:linmap}
\begin{split}
\mathcal{A}_{a, \mu}^{(1)}&=t_a \, \sum_{\ell} \left\{a_{\pm \ell \pm 1} \, \phi^{(1)}_{\ell\pm 1}\, V_{\ell,i} \, \delta_{\mu}^{i} +   \, \frac{1}{r^{d-3}} \epsilon_{\alpha \beta}\hat{\nabla}^{\beta}\left(r^{d-4}\,a_{\pm\ell} \,a_{\pm\ell \pm 2} \, \phi^{(1)}_{\ell\pm 2}\right)  \, S_{\ell} \, \delta_{\mu}^{\alpha} \right\}.\\
\end{split}
\end{equation}
Here, choosing all the plus signs reproduces Eq. (\ref{eq:linmapone}), whereas choosing all the minus signs reproduces Eq. (\ref{eq:linmaptwo}).

As we have mentioned, we can arrive at an analogous map if we choose to lower or raise the frequency $\Delta$ instead. Choosing to lower frequency, we arrive at
\begin{equation}\begin{split} 
\phi^V_{\Delta,\ell}&= \frac{r}{f(r)^{\half}}\left(\left(\Delta+1\right)-r\, \fofr\, \frac{\partial}{\partial r}\right) \phi_{\Delta+1,\ell} \ ,\\
\phi^S_{\Delta,\ell}&= \frac{r}{f(r)^{\half}}\left(\left(\Delta+1\right)-r\, \fofr\, \frac{\partial}{\partial r}\right)\phi^V_{\Delta+1,\ell}\\
&=\frac{r^2}{f(r)}\left(\left(\Delta+1\right)-r\, \fofr\, \frac{\partial}{\partial r}\right)^2 \phi_{\Delta+2,\ell} 
\end{split}\end{equation} 

Defining a \emph{lowering} operator in angular momentum space, $b_{\Delta+1}=\frac{r}{f(r)^{\half}}\left\{\left(\Delta+1\right)-r\, \fofr\, \frac{\partial}{\partial r}\right\}$, we can cast these relations as
\begin{equation}\begin{split} 
\phi^V_{\Delta,\ell}&=b_{\Delta+1} \, \phi_{\Delta+1,\ell} \ ,\\
\phi^S_{\Delta,\ell}&=b_{\Delta+1}\,\phi^V_{\Delta+1,\ell}\\
&=\left(b_{\Delta+1}\right)^2 \, \phi_{\Delta+2,\ell}
\end{split}\end{equation}

If we raise the frequency instead, the corresponding relations are
\begin{equation}\begin{split} 
\phi^V_{\Delta,\ell}&= \frac{r}{f(r)^{\half}}\left(\left(\Delta-1\right)+r\, \fofr\, \frac{\partial}{\partial r}\right) \phi_{\Delta-1,\ell} \ ,\\
\phi^S_{\Delta,\ell}&= \frac{r}{f(r)^{\half}}\left(\left(\Delta-1\right)+r\, \fofr\, \frac{\partial}{\partial r}\right)\phi^V_{\Delta-1,\ell}\\
&=\frac{r^2}{f(r)}\left(\left(\Delta-1\right)+r\, \fofr\, \frac{\partial}{\partial r}\right)^2 \phi_{\Delta-2,\ell} 
\end{split}\end{equation} 

In the same vein as earlier, we can define a \emph{raising} operator in frequency space as
$b_{-\Delta-1}=\frac{r}{f(r)^{\half}}\left\{\left(\Delta-1\right)+r\, \fofr\, \frac{\partial}{\partial r}\right\}$, and express these relations as
\begin{equation}\begin{split} 
\phi^V_{\Delta,\ell}&=b_{-\Delta-1} \, \phi_{\Delta-1,\ell} \ ,\\
\phi^S_{\Delta,\ell}&=b_{-\Delta-1}\,\phi^V_{\Delta-1,\ell}\\
&=\left(b_{-\Delta-1}\right)^2 \, \phi_{\Delta-2,\ell}
\end{split}\end{equation}
As in the case of the $a_k$ operators, we have a definition of $b_k$ for all integer $k$. The negative integers provide the raising operators, and the non-negative integers provide the lowering operators. Using these operators, we see that for every solution of the bi-adjoint equations of motion given by Eq. (\ref{eq:balinsoln}), we can generate a solution to the linearized Yang-Mills equations, analogous to Eq.(\ref{eq:linmap}) given by
\begin{equation} \label{eq:linmapdelta}
\begin{split}
\mathcal{A}_{a, \mu}^{(1)}&=t_a \, \sum_{\ell,\Delta}  \left\{e^{-i \frac{\Delta}{R} t}\,b_{\pm \Delta \pm 1} \, \phi^{(1)}_{\Delta\pm 1,\ell}\, V_{\ell,i} \, \delta_{\mu}^{i} +   \, \frac{1}{r^{d-3}} \epsilon_{\alpha \beta}\hat{\nabla}^{\beta}\left(e^{-i \frac{\Delta}{R} t}r^{d-4}\,(b_{\pm \Delta \pm 1})^2\, \phi^{(1)}_{\Delta\pm 2,\ell}\right)  \, S_{\ell} \, \delta_{\mu}^{\alpha} \right\}
\end{split}
\end{equation}
where, choosing all the plus signs generates solutions by lowering $\Delta$ whereas choosing all the minus signs generates solutions by raising $\Delta$.

\section{The bi-adjoint to Yang-Mills correspondence to all orders}
\label{sec:allordermap}
In this section, we will provide a map between all perturbative solutions of the bi-adjoint equations of motion, and those of the Yang-Mills equations in an arbitrary curved spacetime. 
We shall specialize the all-order map to flat spacetimes in Sec. (\ref{sec:allordermink}), and to AdS spacetimes in Sec.(\ref{sec:allorderAdS}).

The all-order map between the bi-adjoint and Yang-Mills solutions takes as input a linearized map between
these solutions. However, it doesn't depend on the specific form of this map. For notational simplicity we shall refer to the linearized map as 		
\begin{equation} \label{eq:linearmap}
\mathcal{A}_{a, \mu}^{(1)}= t_a \, \mathcal{K}_{\mu}^{(1)} \left[\phi^{(1)}\right]=\mathcal{K}_{a,\mu}^{(1)} \left[\phi^{(1)}\right].
\end{equation}
This map could be the one described in the previous section for maximally symmetric spacetimes as in Eq. (\ref{eq:linmap}). More generally, it could also be some other map between the linearized solutions. For example, one way to generalize the map in the previous section is to introduce non-dynamical sources, whose effect is only seen at the linear order in perturbation. It could also be an entirely different map constructed for a non maximally symmetric spacetime. The all-order map we construct in this section can be used to obtain the respective perturbations in a general curved spacetime, about any such solution. The all-order map is also agnostic to the choice of gauge.

We wish to lift the linear order map in Eq. (\ref{eq:linmap}) to an arbitrary order $n$. As the all-order map can be cast in terms of just the linear bi-adjoint solution, we will abuse notation for simplicity and write the map as
\begin{equation} \label{eq:allordermap}
\mathcal{A}_{a, \mu}^{(n)}= \mathcal{K}_{a,\mu}^{(n)} \left[\phi^{(1)}\right]\equiv \mathcal{K}\phi_{a,\mu}^{(n)} 
\end{equation}
At $n$th order in perturbation, the bi-adjoint equations of motion are
\begin{equation}\label{eq:baeomn}
\nabla^{\mu}\nabla_\mu \phi_{a \tilde{a}}^{(n)}= - \fabc \tfabc \, \sum_{\substack{\{i,j\} \\ \{i+j=n\}}} \phi_{b \tilde{b}}^{(i)}\, \phi_{c \tilde{c}}^{(j)},
\end{equation}
where the sum is over all positive integers $\{i,j\}$ such that $(i+j)=n$. 

We define the bi-adjoint source at order $n$,
\begin{equation}
\mathcal{J}_{\tilde{a},bc}^{(n)}=\tfabc \, \sum_{\substack{\{i,j\} \\ \{i+j=n\}}} \phi_{b \tilde{b}}^{(i)}\, \phi_{c \tilde{c}}^{(j)},
\end{equation}
The solution of the bi-adjoint equations of motion are integrals over the spacetime with the integrand being a convolution of the Green's function with the bi-adjoint source $\mathcal{J}_{\tilde{a},bc}^{(n)}$. 

In the spirit of the quantum double copy procedure, the aim is to provide the analogous integrand for the Yang-Mills field $\mathcal{A}_{a,\mu}^{(n)}$. 
The Yang-Mills equations 
\begin{equation}
\nabla^{\mu} F^{a}_{\mu \nu}= - \fabc \, A_{b}^{\mu}\, F^{c}_{\mu \nu}
\end{equation}
can be cast, to $n$th order in perturbation, as 
\begin{equation}\label{eq:ymnthorder}
\nabla^{\mu} \left(\nabla_{\mu} \mathcal{A}_{a,\nu}^{(n)}-\nabla_{\nu} \mathcal{A}_{a,\mu}^{(n)} \right)=-\fabc \, \mathcal{J}_{\mu,bc}^{(n)}.
\end{equation}
The existence of a map Eq.(\ref{eq:linmap}) between the linearized bi-adjoint solutions and the linearized Yang-Mills solutions implies a map between the corresponding Green's functions. The solution to the gauge field at arbitrary order in perturbation can be cast as an integral of this Green's function convoluted with the Yang-Mills source $\mathcal{J}_{\mu,bc}^{(n)}$.

To specify the all order map, what remains then is to provide the Yang-Mills source $\mathcal{J}_{\mu,bc}^{(n)}$ in terms of the linearized bi-adjoint solutions. In order to so, let us first introduce the operator 
\begin{equation} \label{eq:goperator}
\mathcal{G}^{\mu \nu \rho}[x_1,x_2,x_3] = \half \left\{ g^{\mu \nu} \left(\nabla_2-\nabla_1\right)^{\rho}+g^{\nu \rho} \left(\nabla_3-\nabla_2\right)^{\mu}+g^{\rho \mu} \left(\nabla_1-\nabla_3\right)^{\nu}\right\}
\end{equation}
This is the curved space analogue of the kinematic part of the familiar three-point vertex in pure Yang-Mills theory on a flat background. The action of this operator on a product of fields, say $C_\nu \, C_\rho$ is defined in a point-splitting way. We first obtain the action of the operator on the product by separating the fields in space-time i.e, we first compute $\mathcal{G}^{\mu \nu \rho}[x_1,x_2,x_3] C_\nu(x_2) \, C_\rho(x_3)$. While doing so, we shall use that the operator $(\nabla_1+\nabla_2+\nabla_3)$ acting on the product of fields is zero. We then take the limit of this expression as the points approach each other i.e $x_2 \rightarrow x_3$.  We denote the final object thus obtained simply as $\mathcal{G}^{\mu \nu \rho} C_\nu \, C_\rho$. 

We are now ready to provide the final ingredient that completes the all-order map,
\begin{equation}\label{eq:ymsourcen}
\mathcal{J}^{\mu(n)}_{ b c}=\sum_{\substack{\{i,j\} \\ \{i+j=n\}}}\left(\mathcal{G}^{\mu \nu \rho} \,  \mathcal{K}\phi_{b,\nu}^{(i)}  \, \, \mathcal{K}\phi_{c,\rho}^{(j)} + f^{cde} \,  \mathcal{K}\phi_{b}^{\lambda(i)} \sum_{\substack{\{p,q\} \\ \{p+q=j\}}} \mathcal{K}\phi_{d,\lambda}^{(p)} \,  \mathcal{K}\phi_{e}^{\mu(q)} \right).
\end{equation}

We shall unpack this map now by writing it down explicitly for $n=2,3$. The solutions thus generated are seen to satisfy the Yang-Mills equations by an explicit computation. We then generalize the argument to arbitrary $n$.

At second order in the coupling, the bi-adjoint equation of motion Eq.(\ref{eq:baeomn}) reads
\begin{equation} \label{eq:baeom2}
\nabla^{\mu}\nabla_\mu \phi_{a \tilde{a}}^{(2)}= -\fabc \mathcal{J}^{\ta(2)}_{b c}, 
\end{equation}
with the bi-adjoint source term defined as 
\begin{equation} \label{eq:basource}
\mathcal{J}^{\ta(2)}_{b c}= \tfabc \, \phi_{b \tilde{b}}^{(1)}\, \phi_{c \tilde{c}}^{(1)}.
\end{equation}
In terms of Feynman diagrams, this corresponds to a diagram with a 3-point vertex, with two source insertions corresponding to the first order solution. The operator $\mathcal{G}^{\mu \nu \rho}$ works to reproduce the corresponding Yang-Mills contribution, so that the Yang-Mills source Eq.(\ref{eq:ymsourcen}) at the second order is

\begin{equation}\label{eq:2ndordermap}
\mathcal{J}^{\mu(2)}_{ b c}= \mathcal{G}^{\mu \nu \rho} \, t_b \, \mathcal{K}_{\nu}^{(1)}\left[\phi^{(1)}\right] t_c \, \mathcal{K}_{\rho}^{(1)} \left[\phi^{(1)}\right].
\end{equation}

This provides the second order gauge field entirely in terms of solutions to the bi-adjoint equation of motion. As we have the linear map Eq. (\ref{eq:linmap}), all we need in order to verify the map at the second order is to reproduce the second order Yang-Mills source. From the definition in Eq.(\ref{eq:ymnthorder}), this is
\begin{equation}
\{2 (\nabla^{\lambda}\mathcal{A}^{\mu(1)}_{b}) \mathcal{A}^{(1)}_{c,\lambda}+\mathcal{A}^{\mu(1)}_{b}  (\nabla^{\lambda}\mathcal{A}^{(1)}_{c,\lambda})+\mathcal{A}^{\lambda(1)}_b \nabla^{\mu}\mathcal{A}^{(1)}_{c,\lambda}\}
\end{equation}

Upon using the first order map Eq.(\ref{eq:linmap}), it reduces to the form in Eq.(\ref{eq:2ndordermap}). 

We now proceed to the third order. As mentioned earlier, we abuse notation to write the second order map as
\begin{equation}
\mathcal{A}_{a,\mu}^{(2)}= \mathcal{K}_{a, \mu}^{(2)} \left[\phi^{(1)}\right]=\mathcal{K}\phi_{a, \mu}^{(2)}
\end{equation}

At the third order, the bi-adjoint source is
\begin{equation} \label{eq:basource3}
\mathcal{J}^{\ta(3)}_{b c}= 2 \,\tfabc \, \phi_{b \tilde{b}}^{(1)}\, \phi_{c \tilde{c}}^{(2)}
\end{equation}
The second order solution for the bi-adjoint scalar $\phi_{c \tilde{c}}^{(2)}$ is linear in both structure constants $\fabc$ and $\tfabc$. Hence, the third order solution is quadratic in $\fabc$ and $\tfabc$. This corresponds to Feynman diagrams with two 3-point vertices connected by an internal propagator. The second term in the corresponding Yang-Mills source in Eq. (\ref{eq:ymsourcen}) \emph{contracts} part of this contribution into a 4-point vertex to give
\begin{equation} \label{eq:ymsource3}
\mathcal{J}^{\mu(3)}_{ b c}=2\,\mathcal{G}^{\mu \nu \rho} \,  \mathcal{K}\phi_{b,\nu}^{(1)}  \, \, \mathcal{K}\phi_{c,\rho}^{(2)} +f^{cde} \,  \mathcal{K}\phi_{b}^{\lambda(1)} \,  \mathcal{K}\phi_{d,\lambda}^{(1)} \,  \mathcal{K}\phi_{e}^{\mu(1)} .
\end{equation}
In other words, this is the inverse of the \emph{blowing up} procedure of contributions of contact vertices into those of 3-point vertices that is used in the discussion of the BCJ double copy for quantum amplitudes.
Taken together with the second order map, it is clear that the third order map is again entirely in terms of the linearized bi-adjoint solution. The third order Yang-Mills source is 
\begin{equation}
\begin{split}
&2 \, \{2 (\nabla^{\lambda}\mathcal{A}^{\mu(1)}_{b}) \mathcal{A}^{(2)}_{c,\lambda}+\mathcal{A}^{\mu(1)}_{b}  (\nabla^{\lambda}\mathcal{A}^{(2)}_{c,\lambda})+\mathcal{A}^{\lambda(1)}_b \nabla^{\mu}\mathcal{A}^{(2)}_{c,\lambda}\} -\fabc f^{cde} \mathcal{A}_{b}^{\lambda(1)} \,  \mathcal{A}_{d,\lambda}^{(1)} \,  \mathcal{A}_{e}^{\mu(1)}.
\end{split}
\end{equation}

Using the first and second order map, this can be seen to take the form quoted in Eq.(\ref{eq:ymsource3}).

The generalization of the map to arbitrary order $n$ is immediate. We first cast the Yang-Mills source at order $n$ into the form
\begin{equation}
\begin{split}
&\sum_{\substack{\{i,j\} \\ \{i+j=n\}}} \left\{2 (\nabla^{\lambda}\mathcal{A}^{\mu(i)}_{b}) \mathcal{A}^{(j)}_{c,\lambda}+\mathcal{A}^{\mu(i)}_{b}  (\nabla^{\lambda}\mathcal{A}^{(j)}_{c,\lambda})+\mathcal{A}^{\lambda(i)}_b \nabla^{\mu}\mathcal{A}^{(j)}_{c,\lambda}\right\}\\
& -\fabc f^{cde} \sum_{\substack{\{i,j\} \\ \{i+j=n\}}}\mathcal{A}_{b}^{\lambda(i)}  \sum_{\substack{\{p,q\} \\ \{p+q=j\}}}  \mathcal{A}_{d,\lambda}^{(p)} \,  \mathcal{A}_{e}^{\mu(q)}
\end{split}
\end{equation}

We then use the maps of order smaller than $n$ to express this in the form quoted in Eq.(\ref{eq:ymsourcen}).

\subsection{Flat space}
\label{sec:allordermink}

The all-order map of the last section was between perturbations of the bi-adjoint scalar and those of the Yang-Mills field in an arbitrary spacetime, and without any gauge choice. We specialize this map to Minkowski spacetimes. We first need to specify the linearized map. We could use the flat space limit of the linearized map of Sec.\ref{sec:linmap}, which follows from the recurrence relations satisfied by Bessel functions. However, in flat space, there's a simpler choice for the linearized map that we shall use here. We choose the Coulomb gauge $\nabla^\mu \mathcal{A}_{a, \mu}=0$, and find it convenient to write the solutions in momentum space.

The solution of the bi-adjoint equations of motion 
\begin{equation}
\partial^{\mu}\partial_\mu \phi_{a \tilde{a}}= - \fabc \tfabc \, \phi_{b \tilde{b}}\, \phi_{c \tilde{c}}
\end{equation}
can be written, in momentum space, as
\begin{equation}
\mathcal{\phi}_{a \tilde{a}}(k)= \frac{1}{k^2} \fabc \mathcal{J}^{\ta}_{b c}(k),
\end{equation}
\begin{equation}
\mathcal{J}^{\ta}_{b c}(k)= \tfabc \int_{k_1,k_2} \phi_{b \tilde{b}}(k_1)\, \phi_{c \tilde{c}}(k_2) \delta^{(d)}(k-k_1-k_2),
\end{equation}
where we have introduced the notation $\int_k = \int \frac{d^d k}{(2 \pi)^d} $.

In the Coulomb gauge, the solution of the Yang-Mills equations
\begin{equation}
\partial^{\mu}\partial_\mu A_{a}^{\nu}(x)= - \fabc J^{\nu}_{b c}(x), \\
\end{equation}
\begin{equation}\label{eq:flatymcurrent}
J^{\nu}_{b c}(x)=A^{b}_{\nu} (x)\left( \partial^{\nu}A^{\mu}_c(x) - F^{\mu \nu}_c(x) \right)
\end{equation}
can be similarly written, in momentum space, as
\begin{equation}
\mathcal{A}_{a}^{\nu}(k)= \frac{1}{k^2} \fabc \mathcal{J}^{\nu}_{b c}(k),
\end{equation}
where $\mathcal{J}^{\nu}_{b c}(k)$ is the Fourier transform of Eq.~(\ref{eq:flatymcurrent}).

Starting with a linearized solution of the bi-adjoint equations 
\[\phi_{a \tilde{a}}^{(1)}(k)= t_a t_{\ta} \, \phi^{(1)}(k), \]
it is easily seen that the linearized Yang-Mills equations of motion are solved by \begin{equation}\label{eq:flatym1map}
\mathcal{A}^{(1)}_{a,\nu}(k)= t_a \, k_{\nu} \, \phi^{(1)}(k) \equiv \mathcal{K}\phi_{a, \nu}^{(1)}(k).
\end{equation}
This defines the first order map. We now use the all-order map to arrive at the higher order Yang-Mills solutions. We first define the flat space version of Eq. \ref{eq:goperator} in momentum space,
\begin{equation} \label{eq:goperatormom}
\Gamma^{\mu \nu \rho}(k_1,k_2,k_3) = -\frac{i}{2} \left\{\eta^{\mu \nu} \left(k_2-k_1\right)^{\rho}+\eta^{\nu \rho} \left(k_3-k_2\right)^{\mu}+\eta^{\rho \mu} \left(k_1-k_3\right)^{\nu}\right\} 
\end{equation}

At the second order, using Eq.~(\ref{eq:2ndordermap}), we get
\begin{equation}\label{eq:flatym2map}
\begin{split}
\mathcal{A}_{a}^{\mu(2)}(k)&=\frac{g}{k^2} \fabc \, t_b t_c \int_{k_1,k_2}  \Gamma^{\mu \nu \rho}(-k,k_1,k_2)\,k_{1,\nu} \, k_{2,\rho}\,\phi^{(1)}(k_1) \, \phi^{(1)}(k_2) \,\delta^{(d)}(k-k_1-k_2) \\
&=\frac{g}{k^2} \fabc \, t_b t_c \int_{p}  \Gamma^{\mu \nu \rho}(-k,p,k-p)\,p_{\nu} \, (k-p)_{\rho}\,\phi^{(1)}(p) \, \phi^{(1)}(k-p)  \\
&\equiv \mathcal{K}\phi_{a}^{\mu(2)}(k)
\end{split}
\end{equation}
This can be explicitly checked to be a solution of the second order Yang-Mills equations. \\

At the third order, we use Eq. (\ref{eq:ymsource3}) to construct
\begin{equation}
\begin{split}
\mathcal{J}^{\mu(3)}_{ b c}(k)= &2 \int_{k_1,k_2} \Gamma^{\mu \nu \rho}(-k,k_1,k_2) \,  \mathcal{K}\phi_{b,\nu}^{(1)}(k_1)  \, \, \mathcal{K}\phi_{c,\rho}^{(2)}(k_2) \delta^{(d)}(k-k_1-k_2) \\
&+ f^{cde}\int_{k_1,k_2,k_3}  \mathcal{K}\phi_{b}^{\nu(1)}(k_1) \,  \mathcal{K}\phi_{d,\nu}^{(1)}(k_2) \,  \mathcal{K}\phi_{e}^{\mu(1)}(k_3) \delta^{(d)}(k-k_1-k_2-k_3)
\end{split}
\end{equation}
After substituting the lower order maps found in Eqs.~(\ref{eq:flatym1map},\ref{eq:flatym2map}), we find that the Yang-Mills field is given by
\begin{equation}
\begin{split}
\mathcal{A}_{a}^{\mu(3)}(k)= &\frac{g^2}{k^2}  \fabc f^{cde} t_b t_d t_e \int_{k_1,k_2,k_3} \phi^{(1)}(k_1) \,\phi^{(1)}(k_2) \, \phi^{(1)}(k_3) \delta^{(d)}(k-k_1-k_2-k_3)\\
&\times \left\{ \frac{2}{(k-k_1)^2} \Gamma^{\mu \nu \rho}(-k,k_1,k-k_1)\, \Gamma^{\theta \gamma \sigma}(-(k-k_1),k_2,k_3) \eta_{\theta \rho}+\eta^{\mu \sigma}\eta^{\gamma \nu}\right\} k_{1,\nu} k_{2,\gamma}
k_{3,\sigma}\\
=&\frac{g^2}{k^2}  \fabc f^{cde} t_b t_d t_e \int_{p,q} p_{\nu} q_{\gamma}
(k-p-q)_{\sigma} \, \phi^{(1)}(p) \,\phi^{(1)}(q) \, \phi^{(1)}(k-p-q) \\
&\times \left\{ \frac{2 \, \eta_{\theta \rho}}{(k-p)^2} \Gamma^{\mu \nu \rho}(-k,p,k-p)\, \Gamma^{\theta \gamma \sigma}(-(k-p),q,k-p-q) +\eta^{\mu \sigma}\eta^{\gamma \nu}\right\} 
 \end{split}
\end{equation}

Again, this can be explicitly checked to be a solution to the Yang-Mills equations at this order. 

Continuing this procedure, we can produce explicit expressions for the Yang-Mills field in terms of the linearized bi-adjoint solutions, at an arbitrary order in the perturbation.

\subsection{AdS}
\label{sec:allorderAdS}

In this section, we specialize to $AdS_{d+1}$, and choose the gauge described in Sec. (\ref{sec:ymlin}), in which the Yang Mills field is given by Eqs. (\ref{eq:ymgauge_allorder}). Beginning from solutions to the bi-adjoint equation of motion in $AdS_{d+1}$ at the $n$th order, the map described in the previous section gives us $\mathcal{J}^{\mu(n)}_{ b c}$, defined by Eq. (\ref{eq:ymsourcen}). We decompose this, just as we did for the Yang-Mills field in Eq. (\ref{eq:GaugeDecomp}), to get the vector and scalar components $J^{V(n)}_{a}$ and $J^{S(n)}_{a}$,
\begin{equation}
\begin{split}
\fabc \mathcal{J}^{\mu(n)}_{ b c}&=  J^{V(n)}_{a} V_i \, \delta_{\mu}^{i}+   J^{S(n)}_{a} S \, \delta_{\mu}^{\alpha}  \label{eq:SourceDecomp}
\end{split}
\end{equation}
Note that in general, there is also a $D_i S$ component in this decomposition. However, the gauge choice ensures that it doesn't show up in the Yang-Mills field, and so, we have dropped it. The Maxwell equations can be then be cast as second-order differential equations for $\phi^{S(n)}$ and $\phi^{V(n)}$, as in Eq.(\ref{maxwelleqns}) with $J^{S(n)}_{a}$ and $J^{V(n)}_{a}$ as the respective source terms. The linear map between the homogeneous solutions to these equations described in Sec. \ref{sec:linmap} implies that the Green's function for both these equations can be obtained from the Green's function of the linearized bi-adjoint scalar equation. The arbitrary order perturbative solutions for the scalar variables $\phi^{S(n)}$ and $\phi^{V(n)}$ can then be cast as integrals of these Green's functions convoluted with the corresponding sources $J^{S(n)}_{a}$ and $J^{V(n)}_{a}$.

\section{Discussion}
\label{sec:disc}

We have presented a new version of the classical double copy construction for curved spacetimes. Asymptotically AdS spacetimes are a natural arena of interest for these results, due to the gauge gravity correspondence \footnote{See \cite{Farrow:2018yni} for a study of the double copy in the boundary correlators}. For example, we could study gravitationally bound states in the bulk. Having a handle on the classical solutions would provide a systematic way of computing their energies. These correspond to anomalous dimensions of certain boundary operators. Thus, one could hope to systematize the higher order extensions of the leading order calculation provided in \cite{Fitzpatrick:2014vua,Kaviraj:2015xsa} \footnote{Also, see \cite{Kraus:2020nga} for a recent and different approach to this problem.}.

As remarked earlier, one line of future work is to extend our maps to include dynamical matter. Another line of thought is to investigate if the map between classical solutions of scalars, Yang-Mills and gravity implies relations between other quantities of interest, such as effective actions. As a calculational tool too, it would be useful to cast effective actions involving gauge fields or gravity in terms of scalars, thereby providing a common framework for their analysis. In work presented elsewhere \cite{Ghosh:2020lel}, we find that indeed such a line of thought is useful while studying gauge fields and gravity on AdS blackbrane backgrounds.

A particularly interesting example of our results is the application to four dimensional flat spacetimes. Here, the two scalar variables introduced in Sec. \ref{sec:ymlin} turn out to be the same, and this can be thought of as a consequence of electric-magnetic duality~\footnote{I thank R. Loganayagam for pointing this out}. As the gravity solutions can also be written in terms of these same variables, this offers a particularly simple formulation for computing gravitational perturbations around Minkowski spacetimes. One exciting application of this idea is to the calculation of gravitational waves. We shall present this formulation and its comparison with existing descriptions of gravitational radiation in four dimensional flat spacetimes elsewhere.

\section*{Acknowledgments}
I am immensely grateful to R. Loganayagam for initial collaboration, many valuable discussions, and for encouragement to write up these results. I would like to acknowledge Akhil Sivakumar, Anish Kulkarni and Sourav Pandey for collaboration on related work. It is also my pleasure to thank Alok Laddha and Suvrat Raju for useful discussions. I would like to express my sincere gratitude to the people of India for their support in the development of the basic sciences.

\appendix
\section*{Appendix}

\section{Spherical harmonics of $S^{d-1}$} \label{sec:spharms}
Functions on the sphere can be decomposed into eigenfunctions  of the sphere Laplacian. These eigenfunctions, called \emph{spherical harmonics}, are a complete set of orthonormal functions. Here, we shall use symmetric, trace-free and divergence-free spherical harmonics, that we define below. For a detailed analysis of these, refer, for example to~\cite{Higuchi:1986wu}.

As in the main text, the line element we use is
\begin{equation}
	d\Omega_{d-1}^2=\gamma_{ij}d \theta^i d \theta^j,
\end{equation} with $\gamma_{ij}$ the standard round metric on $S^{d-1}$. We use $D_i$ to denote the covariant derivative on the sphere. Spherical harmonics of rank zero, one and two are referred to as scalar, vector and tensor spherical harmonics, and denoted by $S_{\ell},V_{\ell,i}$ and $T_{\ell,ij}$ respectively. 

The scalar spherical harmonics are defined as the solutions to the equation
\begin{equation}\begin{split} 
\left[D^iD_i +\ell(\ell+d-2) \right] S  &= 0\ ,\\
\end{split}\end{equation}
satisfying the orthonormality condition
\begin{equation}
\int d^{d-1}\Omega \, S_{\ell} \, S_{\ell'} = \delta_{\ell,\ell'},
\end{equation}
and where the integers $\ell \geq 0$. These are generalizations of the  Laplace spherical harmonics on the 2-sphere $Y_{\ell,m}(\theta_1,\theta_2)$, and are denoted analogously by $Y_{\ell_1,\ldots \ell_{d-1}}\left(\theta_1, \ldots \theta_{d-1}\right)$.

Vector fields on the sphere can be decomposed into vector spherical harmonics. Rotations of the sphere, and the Laplacian both map the space of all divergence free vector fields onto itself. The vector spherical harmonics are chosen to satisfy
\begin{equation}\begin{split}  \label{vecharm}
\left[D^j D_j +\ell(\ell+d-2)-1 \right] V_
{\ell,i}  &= 0, \qquad D^i V_{\ell,i}=0\\
\end{split}\end{equation}
and
\begin{equation}
\int d^{d-1}\Omega \, V_{\ell,i} \, V_{\ell',i} = \delta_{\ell,\ell'}.
\end{equation}
Here, the integers $\ell \geq 1$. The number of independent vector spherical harmonics, after accounting for the divergence free condition, is $(d-2)$. So, they are non trivial only when the dimension of the sphere $(d-1)$ is at least two, i.e. when the dimension of the full spacetime $(d+1)$ is at least four. They can be obtained by taking appropriate combinations of covariant derivatives acting on $Y_{\ell_1,\ldots \ell_{d-1}}\left(\theta_1, \ldots \theta_{d-1}\right)$. An identity satisfied by the vector spherical harmonics, useful in the derivation of Eqs.(\ref{maxwelleqns}) is
\begin{equation}
	D^j\left(D_i V_{\ell,j}-D_j V_{\ell,i}\right) = (\ell+1)(\ell+d-3)\ V_{\ell,i}\ . 
\end{equation} 
This follows from using Eq. (\ref{vecharm}) and finding the Ricci tensor for the unit $(d-1)$ sphere to be $R_{ij}=(d-2)\gamma_{ij}$. 

Though we won't be needing them in the main text, for completeness, we note that to decompose tensor fields of higher rank, we would need appropriate tensor spherical harmonics. In order to decompose the gravitational field, we need tensor spherical harmonics of rank two. Sphere rotations and the sphere Laplacian map the space of all symmetric, trace free and divergence free tensor fields onto itself. So, we have the tensor spherical harmonics satisfy
\begin{equation}\begin{split}  \label{tenharm}
\left[D^k D_k +\ell(\ell+d-2)-2 \right] T_{\ell,ij}  &= 0, \qquad D^i T_{\ell,ij}=0, \qquad \gamma^{ij}T_{\ell,ij}=0\\
\end{split}\end{equation}
and the orthonormality condition
\begin{equation}
\int d^{d-1}\Omega \, T_{\ell,ij} \, T_{\ell',ij} = \delta_{\ell,\ell'},
\end{equation}
with the integers $\ell \geq 2$. Accounting for the trace-free and divergence-free conditions, the number of such independent symmetric tensor harmonics can be seen to be $(d-1)d/2 -(d-1)-1=d(d-3)/2 $. They are non-trivial only for spheres of dimension at least three, i.e. when the dimension of the full spacetime $(d+1)$ is at least five.
Explicit expressions for the scalar, vector and tensor harmonics can be found in \cite{Higuchi:1986wu}.

\section{Generalized Field Equation} \label{sec:geneq}
In the background given by the $M_{2} \times S^{d-1}$ decomposition of $(d+1)$ dimensional AdS or dS spacetime
\bea
ds^2&=&g_{\mu \nu} \, dx^{\mu} dx^{\nu}= g_{\alpha \beta} \, dy^{\alpha} dy^{\beta} +r^2(y)  d\Omega_{d-1}^2,
\eea
the linearized bi-adjoint, Yang-Mills and Einstein equations can be reduced to hypergeometric differential equations on $M_2$. The generalized form of these equations can be written as
\bea \label{eq:GEq}
D_{\alpha} D^{\alpha} \Phi = \left( \frac{a'}{R^2} + \frac{b'}{r^2} \right) \Phi 
\eea
where $a'$ and $b'$ are constants that parametrize the various solutions (values are given below), and $D_{\alpha}$ is the covariant derivative with respect to the two dimensional metric $g_{\alpha \beta}$. 

We now specialize to the global AdS or dS metric
\bea
ds^2&=&-f(r) dt^2 +\frac{dr^2}{f(r)} + r^2 d\Omega_{d-1}^2, 
\eea
where $f(r)=1\pm\frac{r^2}{R^2}$.

One of the independent solutions to Eq. \ref{eq:GEq}  is
\begin{equation} \label{massol1}
\Phi_1 (t,r)=\sum_{\Delta,\ell} e^{-i \frac{\Delta}{R} t }\,  r^{\frac{1}{2} (1+b)} \left(1+\frac{r^2}{R^2}\right)^{-\frac{\Delta}{2}} \, _2F_1\left(m_{\Delta,-a,b},m_{\Delta,a,b};\frac{2+b}{2};-\frac{r^2}{R^2}\right),  \\
\end{equation} 
where $a=\sqrt{4a'+1}$, $b=\sqrt{4b'+1}$ and $m_{\Delta,a,b}=\frac{1}{4} \left(-2 \Delta +a+b+2\right)$. \\

The other independent solution is related by $b \rightarrow -b$ 
\begin{equation}\label{massol2}
\Phi_2 (t,r)= \sum_{\Delta,\ell} e^{-i \frac{\Delta}{R} t } \, r^{\frac{1}{2} (1-b)} \left(1+\frac{r^2}{R^2}\right)^{-\frac{\Delta}{2}} \, _2F_1\left(m_{\Delta,-a,-b},m_{\Delta,a,-b};\frac{ 2-b}{2};-\frac{r^2}{R^2}\right) \\
\end{equation}

We use $\Phi_{\Delta,\ell}$  to denote the solutions in the frequency space, so that either of the solutions above can be written as $\Phi=\Sigma_{\Delta,\ell} \Phi_{\Delta,\ell}$. 

In the main text, we have shown that the solutions to the linearized equations of motion of the scalar and the gauge field can be written in terms of scalar variables. These were referred to as $\phi_{\Delta,\ell}(t,r)$, $\phi^V_{\Delta,\ell}(t,r)$ and $\phi^S_{\Delta,\ell}(t,r)$. It can be shown that the solutions to the linearized equations of gravity can also be expressed in terms of these same scalar variables. The solution to the linearized bi-adjoint equation of motion and the tensor sector of gravity can be written in terms of $\phi_{\Delta,\ell}(t,r)$. The solutions to the linearized gauge and gravity fields in the vector and scalar sectors can be cast as functions of $\phi^V_{\Delta,\ell}(t,r)$ and $\phi^S_{\Delta,\ell}(t,r)$ respectively. Each of these scalar variables can be expressed in terms of the solutions presented in this section as $r^{- \half c} \, \Phi_{\Delta,\ell}$, with the values of the different parameters tabulated below. \\

\begin{center}
	\begin{tabular}{|c|c|c|c|}
		\hline
		& Bi-adjoint, & Yang-Mills:Vector,& Yang-Mills:Scalar,\\
		&Gravity:Tensor & Gravity:Vector & Gravity:Scalar \\ \hline
		$a$ & $d$ & $d-2$ & $d-4$ \\ 
		\hline 
		$b$ & $d+2\ell-2$ & $d+2\ell-2$ & $d+2\ell-2$ \\
		\hline
		$c$ & $d-1$ & $d-3$ & $d-5$ \\
		\hline
	\end{tabular} 
\end{center}
We note that the parameter $c$ only appears as the exponent of an overall power of $r$, so it can be eliminated by a different choice of the scalar functions. The parameter $b$ is the same for all the solutions, so the different solutions can all be characterized by a single parameter.
\subsection{Asymptotics}
Let us look at the asymptotics of these solutions, as we go to the boundary $r\rightarrow\infty$. We consider the full solution $\Phi(t,r)=c_1 \Phi_1 (t,r)+c_2 \Phi_2 (t,r)$  
\bea
\Phi (t,r) \rightarrow c_1' \, r^{\frac{1}{2} \left(a-1 \right)} + c_2' \, r^{-\frac{1}{2} \left(a+1 \right)},
\eea
where 
\begin{equation}
\begin{split}
c_1'=\frac{\Gamma \left(\frac{a}{2}\right)}{R^{\frac{1}{2} (a-1)}} &\left(\frac{c_1}{R^{\frac{1}{2} (1+b)}}\frac{\Gamma \left(\frac{2+b}{2}\right)}{\Gamma (c_{\Delta,a,b}) \Gamma \left(c_{-\Delta,a,b}\right)}  + \frac{c_2}{R^{\frac{1}{2} (1-b)}} \frac{\Gamma \left(\frac{2-b}{2}\right)}{\Gamma \left(c_{\Delta,a,-b} \right) \Gamma \left(c_{-\Delta,a,-b} \right)} \right) \\
c_2'=\frac{\Gamma \left(-\frac{a}{2}\right)}{R^{-\frac{1}{2} (a+1)}}&\left(\frac{c_1}{R^{\frac{1}{2} (1+b)}}\frac{ \Gamma \left(\frac{2+b}{2}\right)}{\Gamma (c_{\Delta,-a,b}) \Gamma \left(c_{-\Delta,-a,b}\right)}  + \frac{c_2}{R^{\frac{1}{2} (1-b)}} \frac{\Gamma \left(\frac{2-b}{2}\right)}{\Gamma \left(c_{\Delta,-a,-b} \right) \Gamma \left(c_{-\Delta,-a,-b} \right)} \right)  
\end{split}
\end{equation}

The leading radial dependence of the solution near spatial infinity only depends on $a$. This means that, unlike the solution to scalar wave equations in flat spacetime, here the dependence is only on the dimension of spacetime, and not on $\ell$. In other words, all the multipole moments die down at the same rate asymptotically and so, the asymptotic field captures all the multipole moments. 

In contrast, the leading radial behaviour near the origin is only controlled by $b$. So, for all the different cases, we get the same leading behaviour at the origin,
\bea
\Phi (t,r) \rightarrow c_1 \, \left(\frac{r}{R}\right)^{\frac{1}{2}(d+2l-1)}+ c_2 \, \left(\frac{r}{R}\right)^{-\frac{1}{2}(d+2l+1)} 
\eea
The choice of $c_1=0$ ensures regularity at the origin. 

\subsection{Flat space limit}
We shall now consider the flat space limit of the solutions to the generalized field equation Eq. \ref{eq:GEq}. We express the first of these solutions in a suitable form, 
\begin{equation}
\Phi_1 (t,r)=\sum_{\omega,\ell} e^{-i \omega t }\,  r^{\frac{1}{2} (1+b)} \left(1\pm\frac{r^2}{R^2}\right)^{-\frac{\omega R}{2}} \, _2F_1\left(m_{\omega R,-a,b},m_{\omega R,a,b};\frac{2+b}{2};-\frac{r^2}{R^2}\right), 
\end{equation}
where we have identified the frequency as $\omega=\frac{\Delta}{R}$, and $a,b$ take values as specified in the last section.
The flat space limit is obtained by taking $R$ to infinity,
\begin{equation}
\begin{split}
&\lim_{R \to \infty}\Phi_1  (t,r)\\
&=\lim_{R \to \infty}\sum_{\omega,\ell}  \left\{e^{-i \omega t }\,   r^{\frac{1}{2} (1+b)} \left(1\pm\frac{r^2}{R^2}\right)^{-\frac{\omega R}{2}}  \right.\\
& \left.   \times \,  _2F_1\left(\frac{1}{4} \left(-2 \omega R -a+b+2\right),\frac{1}{4} \left(-2 \omega R +a+b+2\right);\frac{2+b}{2};-\frac{\frac{\omega^2 r^2}{4}}{\frac{1}{16} \times \left((b+2-2\omega R)^2-a^2\right) }\right) \right\}
\end{split}
\end{equation}
In the flat space limit, two of the regular singularities of the hypergeometric functions merge, and we get a confluent hypergeometric function. Using the formula
\begin{equation}\begin{split} 
\lim_{a,b\to \infty}\ {}_2F_1\left[a,b ; c ; \frac{x}{ab}\right] ={}_0F_1\left[ c ; x\right]  \  \end{split}\end{equation}
yields
\begin{equation}
\begin{split}
&\lim_{R \to \infty}\Phi_1 (t,r)
=\sum_{\omega,\ell} e^{-i \omega t }\,   r^{\frac{1}{2} (1+b)} \, _0F_1\left(\frac{2+b}{2};-\frac{\omega^2 r^2}{4}\right). 
\end{split}
\end{equation}
This is seen to be a Bessel function, as expected, via
\begin{equation}\begin{split} 
J_\nu(x)
&=  \frac{\left(\frac{x}{2}\right)^\nu}{\Gamma\left(\nu+1\right)} {}_0F_1\left[\nu+1 ; -\frac{x^2}{4}\right]\ , \end{split}\end{equation}
which leads to
\begin{equation}\begin{split} \label{eq:flat1}
\lim_{R\to\infty}  \Phi_1 (t,r) &=\sum_{\omega,\ell}    \frac{\Gamma(\frac{2+b}{2}) }{
	\left(\frac{\omega}{2}\right)^{\frac{b}{2}}} \ e^{-i \omega t } \, \sqrt{r} \, J_{\frac{b}{2}}(\omega r)\ . \end{split}\end{equation}
Similarly for the other solution in Eq.\ref{massol2}, we find 
\begin{equation}\begin{split} 
\lim_{R\to\infty}  \Phi_2 (t,r) 
&=\sum_{\omega,\ell}  \left(\frac{\omega}{2}\right)^{\frac{b}{2}}  \Gamma\left(\frac{2-b}{2}\right) \ e^{-i \omega t } \, \sqrt{r} \,J_{-\frac{b}{2}}(\omega r)\ , 
\end{split}\end{equation}
where, as mentioned before, $b=d+2\ell-2$.

The flat space limit of the linearized map in Sec.\ref{sec:linmap} then follows from the recurrence relations satisfied by Bessel functions.

\providecommand{\href}[2]{#2}\begingroup\raggedright
\endgroup


\begin{thebibliography}{10}
	
	\bibitem{Kawai:1985xq}
	H.~Kawai, D.C.~Lewellen and S.H.H.~Tye, \emph{{A Relation Between Tree
			Amplitudes of Closed and Open Strings}},
	\href{https://doi.org/10.1016/0550-3213(86)90362-7}{\emph{Nucl. Phys.}
		{\bfseries B269} (1986) 1}.
	
	\bibitem{Bern:2008qj}
	Z.~Bern, J.J.M.~Carrasco and H.~Johansson, \emph{{New Relations for
			Gauge-Theory Amplitudes}},
	\href{https://doi.org/10.1103/PhysRevD.78.085011}{\emph{Phys. Rev.}
		{\bfseries D78} (2008) 085011}
	[\href{https://arxiv.org/abs/0805.3993}{{\ttfamily 0805.3993}}].
	
	\bibitem{Bern:2010ue}
	Z.~Bern, J.J.M.~Carrasco and H.~Johansson, \emph{{Perturbative Quantum Gravity
			as a Double Copy of Gauge Theory}},
	\href{https://doi.org/10.1103/PhysRevLett.105.061602}{\emph{Phys. Rev. Lett.}
		{\bfseries 105} (2010) 061602}
	[\href{https://arxiv.org/abs/1004.0476}{{\ttfamily 1004.0476}}].
	
	\bibitem{Bern:2010yg}
	Z.~Bern, T.~Dennen, Y.-t.~Huang and M.~Kiermaier, \emph{{Gravity as the Square
			of Gauge Theory}},
	\href{https://doi.org/10.1103/PhysRevD.82.065003}{\emph{Phys. Rev. D}
		{\bfseries 82} (2010) 065003}
	[\href{https://arxiv.org/abs/1004.0693}{{\ttfamily 1004.0693}}].
	
	\bibitem{BjerrumBohr:2010hn}
	N.E.J.~Bjerrum-Bohr, P.H.~Damgaard, T.~Sondergaard and P.~Vanhove, \emph{{The
			Momentum Kernel of Gauge and Gravity Theories}},
	\href{https://doi.org/10.1007/JHEP01(2011)001}{\emph{JHEP} {\bfseries 01}
		(2011) 001} [\href{https://arxiv.org/abs/1010.3933}{{\ttfamily 1010.3933}}].
	
	\bibitem{Mafra:2011kj}
	C.R.~Mafra, O.~Schlotterer and S.~Stieberger, \emph{{Explicit BCJ Numerators
			from Pure Spinors}},
	\href{https://doi.org/10.1007/JHEP07(2011)092}{\emph{JHEP} {\bfseries 07}
		(2011) 092} [\href{https://arxiv.org/abs/1104.5224}{{\ttfamily 1104.5224}}].
	
	\bibitem{Bjerrum-Bohr:2016axv}
	N.E.J.~Bjerrum-Bohr, J.L.~Bourjaily, P.H.~Damgaard and B.~Feng,
	\emph{{Manifesting Color-Kinematics Duality in the Scattering Equation
			Formalism}}, \href{https://doi.org/10.1007/JHEP09(2016)094}{\emph{JHEP}
		{\bfseries 09} (2016) 094}
	[\href{https://arxiv.org/abs/1608.00006}{{\ttfamily 1608.00006}}].
	
	\bibitem{Bern:2017ucb}
	Z.~Bern, J.J.M.~Carrasco, W.-M.~Chen, H.~Johansson, R.~Roiban and M.~Zeng,
	\emph{{Five-loop four-point integrand of $N=8$ supergravity as a generalized
			double copy}}, \href{https://doi.org/10.1103/PhysRevD.96.126012}{\emph{Phys.
			Rev.} {\bfseries D96} (2017) 126012}
	[\href{https://arxiv.org/abs/1708.06807}{{\ttfamily 1708.06807}}].
	
	\bibitem{Bern:2018jmv}
	Z.~Bern, J.J.~Carrasco, W.-M.~Chen, A.~Edison, H.~Johansson, J.~Parra-Martinez
	et~al., \emph{{Ultraviolet Properties of $\mathcal N = 8$ Supergravity at
			Five Loops}}, \href{https://doi.org/10.1103/PhysRevD.98.086021}{\emph{Phys.
			Rev. D} {\bfseries 98} (2018) 086021}
	[\href{https://arxiv.org/abs/1804.09311}{{\ttfamily 1804.09311}}].
	
	\bibitem{Carrasco:2015iwa}
	J.J.M.~Carrasco, \emph{{Gauge and Gravity Amplitude Relations}},  in
	\emph{{Proceedings, Theoretical Advanced Study Institute in Elementary
			Particle Physics: Journeys Through the Precision Frontier: Amplitudes for
			Colliders (TASI 2014): Boulder, Colorado, June 2-27, 2014}}, pp.~477--557,
	WSP, WSP, 2015, \href{https://doi.org/10.1142/9789814678766_0011}{DOI}
	[\href{https://arxiv.org/abs/1506.00974}{{\ttfamily 1506.00974}}].
	
	\bibitem{Borsten:2020bgv}
	L.~Borsten, \emph{{Gravity as the square of gauge theory: a review}},
	\href{https://doi.org/10.1007/s40766-020-00003-6}{\emph{Riv. Nuovo Cim.}
		{\bfseries 43} (2020) 97}.
	
	\bibitem{Bern:2019prr}
	Z.~Bern, J.J.~Carrasco, M.~Chiodaroli, H.~Johansson and R.~Roiban, \emph{{The
			Duality Between Color and Kinematics and its Applications}},
	\href{https://arxiv.org/abs/1909.01358}{{\ttfamily 1909.01358}}.
	
	\bibitem{Monteiro:2014cda}
	R.~Monteiro, D.~O'Connell and C.D.~White, \emph{{Black holes and the double
			copy}}, \href{https://doi.org/10.1007/JHEP12(2014)056}{\emph{JHEP} {\bfseries
			12} (2014) 056} [\href{https://arxiv.org/abs/1410.0239}{{\ttfamily
			1410.0239}}].
	
	\bibitem{Luna:2015paa}
	A.~Luna, R.~Monteiro, D.~O'Connell and C.D.~White, \emph{{The classical double
			copy for Taub–NUT spacetime}},
	\href{https://doi.org/10.1016/j.physletb.2015.09.021}{\emph{Phys. Lett.}
		{\bfseries B750} (2015) 272}
	[\href{https://arxiv.org/abs/1507.01869}{{\ttfamily 1507.01869}}].
	
	\bibitem{Luna:2016due}
	A.~Luna, R.~Monteiro, I.~Nicholson, D.~O'Connell and C.D.~White, \emph{{The
			double copy: Bremsstrahlung and accelerating black holes}},
	\href{https://doi.org/10.1007/JHEP06(2016)023}{\emph{JHEP} {\bfseries 06}
		(2016) 023} [\href{https://arxiv.org/abs/1603.05737}{{\ttfamily
			1603.05737}}].
	
	\bibitem{Luna:2016hge}
	A.~Luna, R.~Monteiro, I.~Nicholson, A.~Ochirov, D.~O'Connell, N.~Westerberg
	et~al., \emph{{Perturbative spacetimes from Yang-Mills theory}},
	\href{https://doi.org/10.1007/JHEP04(2017)069}{\emph{JHEP} {\bfseries 04}
		(2017) 069} [\href{https://arxiv.org/abs/1611.07508}{{\ttfamily
			1611.07508}}].
	
	\bibitem{Luna:2017dtq}
	A.~Luna, I.~Nicholson, D.~O'Connell and C.D.~White, \emph{{Inelastic Black Hole
			Scattering from Charged Scalar Amplitudes}},
	\href{https://doi.org/10.1007/JHEP03(2018)044}{\emph{JHEP} {\bfseries 03}
		(2018) 044} [\href{https://arxiv.org/abs/1711.03901}{{\ttfamily
			1711.03901}}].
	
	\bibitem{Bahjat-Abbas:2017htu}
	N.~Bahjat-Abbas, A.~Luna and C.D.~White, \emph{{The Kerr-Schild double copy in
			curved spacetime}},
	\href{https://doi.org/10.1007/JHEP12(2017)004}{\emph{JHEP} {\bfseries 12}
		(2017) 004} [\href{https://arxiv.org/abs/1710.01953}{{\ttfamily
			1710.01953}}].
	
	\bibitem{Carrillo-Gonzalez:2017iyj}
	M.~Carrillo-Gonzalez, R.~Penco and M.~Trodden, \emph{{The classical double copy
			in maximally symmetric spacetimes}},
	\href{https://arxiv.org/abs/1711.01296}{{\ttfamily 1711.01296}}.
	
	\bibitem{Berman:2018hwd}
	D.S.~Berman, E.~Chacón, A.~Luna and C.D.~White, \emph{{The self-dual classical
			double copy, and the Eguchi-Hanson instanton}},
	\href{https://doi.org/10.1007/JHEP01(2019)107}{\emph{JHEP} {\bfseries 01}
		(2019) 107} [\href{https://arxiv.org/abs/1809.04063}{{\ttfamily
			1809.04063}}].
	
	\bibitem{Lee:2018gxc}
	K.~Lee, \emph{{Kerr-Schild Double Field Theory and Classical Double Copy}},
	\href{https://doi.org/10.1007/JHEP10(2018)027}{\emph{JHEP} {\bfseries 10}
		(2018) 027} [\href{https://arxiv.org/abs/1807.08443}{{\ttfamily
			1807.08443}}].
	
	\bibitem{Gurses:2018ckx}
	M.~Gurses and B.~Tekin, \emph{{Classical Double Copy: Kerr-Schild-Kundt metrics
			from Yang-Mills Theory}},
	\href{https://doi.org/10.1103/PhysRevD.98.126017}{\emph{Phys. Rev.}
		{\bfseries D98} (2018) 126017}
	[\href{https://arxiv.org/abs/1810.03411}{{\ttfamily 1810.03411}}].
	
	\bibitem{CarrilloGonzalez:2019gof}
	M.~Carrillo~González, B.~Melcher, K.~Ratliff, S.~Watson and C.D.~White,
	\emph{{The classical double copy in three spacetime dimensions}},
	\href{https://doi.org/10.1007/JHEP07(2019)167}{\emph{JHEP} {\bfseries 07}
		(2019) 167} [\href{https://arxiv.org/abs/1904.11001}{{\ttfamily
			1904.11001}}].
	
	\bibitem{Andrzejewski:2019hub}
	K.~Andrzejewski and S.~Prencel, \emph{{From polarized gravitational waves to
			analytically solvable electromagnetic beams}},
	\href{https://doi.org/10.1103/PhysRevD.100.045006}{\emph{Phys. Rev.}
		{\bfseries D100} (2019) 045006}
	[\href{https://arxiv.org/abs/1901.05255}{{\ttfamily 1901.05255}}].
	
	\bibitem{Cho:2019ype}
	W.~Cho and K.~Lee, \emph{{Heterotic Kerr-Schild Double Field Theory and
			Classical Double Copy}},
	\href{https://doi.org/10.1007/JHEP07(2019)030}{\emph{JHEP} {\bfseries 07}
		(2019) 030} [\href{https://arxiv.org/abs/1904.11650}{{\ttfamily
			1904.11650}}].
	
	\bibitem{Alawadhi:2019urr}
	R.~Alawadhi, D.S.~Berman, B.~Spence and D.~Peinador~Veiga, \emph{{S-duality and
			the double copy}}, \href{https://doi.org/10.1007/JHEP03(2020)059}{\emph{JHEP}
		{\bfseries 03} (2020) 059}
	[\href{https://arxiv.org/abs/1911.06797}{{\ttfamily 1911.06797}}].
	
	\bibitem{Kim:2019jwm}
	K.~Kim, K.~Lee, R.~Monteiro, I.~Nicholson and D.~Peinador~Veiga, \emph{{The
			Classical Double Copy of a Point Charge}},
	\href{https://doi.org/10.1007/JHEP02(2020)046}{\emph{JHEP} {\bfseries 02}
		(2020) 046} [\href{https://arxiv.org/abs/1912.02177}{{\ttfamily
			1912.02177}}].
	
	\bibitem{Bah:2019sda}
	I.~Bah, R.~Dempsey and P.~Weck, \emph{{Kerr-Schild Double Copy and Complex
			Worldlines}}, \href{https://doi.org/10.1007/JHEP02(2020)180}{\emph{JHEP}
		{\bfseries 02} (2020) 180}
	[\href{https://arxiv.org/abs/1910.04197}{{\ttfamily 1910.04197}}].
	
	\bibitem{Arkani-Hamed:2019ymq}
	N.~Arkani-Hamed, Y.-t.~Huang and D.~O'Connell, \emph{{Kerr black holes as
			elementary particles}},
	\href{https://doi.org/10.1007/JHEP01(2020)046}{\emph{JHEP} {\bfseries 01}
		(2020) 046} [\href{https://arxiv.org/abs/1906.10100}{{\ttfamily
			1906.10100}}].
	
	\bibitem{Lescano:2020nve}
	E.~Lescano and J.A.~Rodr\'\i{}guez, \emph{{$ \mathcal{N} $ = 1 supersymmetric
			Double Field Theory and the generalized Kerr-Schild ansatz}},
	\href{https://doi.org/10.1007/JHEP10(2020)148}{\emph{JHEP} {\bfseries 10}
		(2020) 148} [\href{https://arxiv.org/abs/2002.07751}{{\ttfamily
			2002.07751}}].
	
	\bibitem{Elor:2020nqe}
	G.~Elor, K.~Farnsworth, M.L.~Graesser and G.~Herczeg, \emph{{The Newman-Penrose
			Map and the Classical Double Copy}},
	\href{https://doi.org/10.1007/JHEP12(2020)121}{\emph{JHEP} {\bfseries 12}
		(2020) 121} [\href{https://arxiv.org/abs/2006.08630}{{\ttfamily
			2006.08630}}].
	
	\bibitem{Luna:2020adi}
	A.~Luna, S.~Nagy and C.~White, \emph{{The convolutional double copy: a case
			study with a point}},
	\href{https://doi.org/10.1007/JHEP09(2020)062}{\emph{JHEP} {\bfseries 09}
		(2020) 062} [\href{https://arxiv.org/abs/2004.11254}{{\ttfamily
			2004.11254}}].
	
	\bibitem{Easson:2020esh}
	D.A.~Easson, C.~Keeler and T.~Manton, \emph{{Classical double copy of
			nonsingular black holes}},
	\href{https://doi.org/10.1103/PhysRevD.102.086015}{\emph{Phys. Rev. D}
		{\bfseries 102} (2020) 086015}
	[\href{https://arxiv.org/abs/2007.16186}{{\ttfamily 2007.16186}}].
	
	\bibitem{Moynihan:2020ejh}
	N.~Moynihan, \emph{{Scattering Amplitudes and the Double Copy in Topologically
			Massive Theories}},
	\href{https://doi.org/10.1007/JHEP12(2020)163}{\emph{JHEP} {\bfseries 12}
		(2020) 163} [\href{https://arxiv.org/abs/2006.15957}{{\ttfamily
			2006.15957}}].
	
	\bibitem{Alfonsi:2020lub}
	L.~Alfonsi, C.D.~White and S.~Wikeley, \emph{{Topology and Wilson lines: global
			aspects of the double copy}},
	\href{https://doi.org/10.1007/JHEP07(2020)091}{\emph{JHEP} {\bfseries 07}
		(2020) 091} [\href{https://arxiv.org/abs/2004.07181}{{\ttfamily
			2004.07181}}].
	
	\bibitem{Gumus:2020hbb}
	M.K.~Gumus and G.~Alkac, \emph{{More on the classical double copy in three
			spacetime dimensions}},
	\href{https://doi.org/10.1103/PhysRevD.102.024074}{\emph{Phys. Rev. D}
		{\bfseries 102} (2020) 024074}
	[\href{https://arxiv.org/abs/2006.00552}{{\ttfamily 2006.00552}}].
	
	\bibitem{Keeler:2020rcv}
	C.~Keeler, T.~Manton and N.~Monga, \emph{{From Navier-Stokes to Maxwell via
			Einstein}}, \href{https://doi.org/10.1007/JHEP08(2020)147}{\emph{JHEP}
		{\bfseries 08} (2020) 147}
	[\href{https://arxiv.org/abs/2005.04242}{{\ttfamily 2005.04242}}].
	
	\bibitem{Bahjat-Abbas:2020cyb}
	N.~Bahjat-Abbas, R.~Stark-Much\~ao and C.D.~White, \emph{{Monopoles, shockwaves
			and the classical double copy}},
	\href{https://doi.org/10.1007/JHEP04(2020)102}{\emph{JHEP} {\bfseries 04}
		(2020) 102} [\href{https://arxiv.org/abs/2001.09918}{{\ttfamily
			2001.09918}}].
	
	\bibitem{Goldberger:2016iau}
	W.D.~Goldberger and A.K.~Ridgway, \emph{{Radiation and the classical double
			copy for color charges}},
	\href{https://doi.org/10.1103/PhysRevD.95.125010}{\emph{Phys. Rev.}
		{\bfseries D95} (2017) 125010}
	[\href{https://arxiv.org/abs/1611.03493}{{\ttfamily 1611.03493}}].
	
	\bibitem{Goldberger:2017frp}
	W.D.~Goldberger, S.G.~Prabhu and J.O.~Thompson, \emph{{Classical gluon and
			graviton radiation from the bi-adjoint scalar double copy}},
	\href{https://doi.org/10.1103/PhysRevD.96.065009}{\emph{Phys. Rev.}
		{\bfseries D96} (2017) 065009}
	[\href{https://arxiv.org/abs/1705.09263}{{\ttfamily 1705.09263}}].
	
	\bibitem{Goldberger:2017vcg}
	W.D.~Goldberger and A.K.~Ridgway, \emph{{Bound states and the classical double
			copy}}, \href{https://doi.org/10.1103/PhysRevD.97.085019}{\emph{Phys. Rev. D}
		{\bfseries 97} (2018) 085019}
	[\href{https://arxiv.org/abs/1711.09493}{{\ttfamily 1711.09493}}].
	
	\bibitem{Goldberger:2017ogt}
	W.D.~Goldberger, J.~Li and S.G.~Prabhu, \emph{{Spinning particles, axion
			radiation, and the classical double copy}},
	\href{https://doi.org/10.1103/PhysRevD.97.105018}{\emph{Phys. Rev.}
		{\bfseries D97} (2018) 105018}
	[\href{https://arxiv.org/abs/1712.09250}{{\ttfamily 1712.09250}}].
	
	\bibitem{Chester:2017vcz}
	D.~Chester, \emph{{Radiative double copy for Einstein-Yang-Mills theory}},
	\href{https://doi.org/10.1103/PhysRevD.97.084025}{\emph{Phys. Rev. D}
		{\bfseries 97} (2018) 084025}
	[\href{https://arxiv.org/abs/1712.08684}{{\ttfamily 1712.08684}}].
	
	\bibitem{Li:2018qap}
	J.~Li and S.G.~Prabhu, \emph{{Gravitational radiation from the classical
			spinning double copy}},
	\href{https://doi.org/10.1103/PhysRevD.97.105019}{\emph{Phys. Rev.}
		{\bfseries D97} (2018) 105019}
	[\href{https://arxiv.org/abs/1803.02405}{{\ttfamily 1803.02405}}].
	
	\bibitem{Shen:2018ebu}
	C.-H.~Shen, \emph{{Gravitational Radiation from Color-Kinematics Duality}},
	\href{https://doi.org/10.1007/JHEP11(2018)162}{\emph{JHEP} {\bfseries 11}
		(2018) 162} [\href{https://arxiv.org/abs/1806.07388}{{\ttfamily
			1806.07388}}].
	
	\bibitem{Carrillo-Gonzalez:2018pjk}
	M.~Carrillo~González, R.~Penco and M.~Trodden, \emph{{Radiation of scalar
			modes and the classical double copy}},
	\href{https://doi.org/10.1007/JHEP11(2018)065}{\emph{JHEP} {\bfseries 11}
		(2018) 065} [\href{https://arxiv.org/abs/1809.04611}{{\ttfamily
			1809.04611}}].
		
	\bibitem{Carrillo-Gonzalez:2019aao}
		M.~Carrillo~González, R.~Penco, and M.~Trodden, \emph{{Shift symmetries, soft
			limits, and the double copy beyond leading order}}, 	\href{https://doi.org/10.1103/PhysRevD.102.105011}{\emph{Phys. Rev.}
			{\bfseries D102} (2020) 105011}
		[\href{https://arxiv.org/abs/1908.07531}{{\ttfamily 1908.07531}}]
	
	\bibitem{Goldberger:2019xef}
	W.D.~Goldberger and J.~Li, \emph{{Strings, extended objects, and the classical
			double copy}}, \href{https://doi.org/10.1007/JHEP02(2020)092}{\emph{JHEP}
		{\bfseries 02} (2020) 092}
	[\href{https://arxiv.org/abs/1912.01650}{{\ttfamily 1912.01650}}].
	
	\bibitem{Kosower:2018adc}
	D.A.~Kosower, B.~Maybee and D.~O'Connell, \emph{{Amplitudes, Observables, and
			Classical Scattering}},
	\href{https://doi.org/10.1007/JHEP02(2019)137}{\emph{JHEP} {\bfseries 02}
		(2019) 137} [\href{https://arxiv.org/abs/1811.10950}{{\ttfamily
			1811.10950}}].
	
	\bibitem{Maybee:2019jus}
	B.~Maybee, D.~O'Connell and J.~Vines, \emph{{Observables and amplitudes for
			spinning particles and black holes}},
	\href{https://doi.org/10.1007/JHEP12(2019)156}{\emph{JHEP} {\bfseries 12}
		(2019) 156} [\href{https://arxiv.org/abs/1906.09260}{{\ttfamily
			1906.09260}}].
	
	\bibitem{delaCruz:2020bbn}
	L.~de~la Cruz, B.~Maybee, D.~O'Connell and A.~Ross, \emph{{Classical Yang-Mills
			observables from amplitudes}},
	\href{https://doi.org/10.1007/JHEP12(2020)076}{\emph{JHEP} {\bfseries 12}
		(2020) 076} [\href{https://arxiv.org/abs/2009.03842}{{\ttfamily
			2009.03842}}].
	
	\bibitem{Johansson:2019dnu}
	H.~Johansson and A.~Ochirov, \emph{{Double copy for massive quantum particles
			with spin}}, \href{https://doi.org/10.1007/JHEP09(2019)040}{\emph{JHEP}
		{\bfseries 09} (2019) 040}
	[\href{https://arxiv.org/abs/1906.12292}{{\ttfamily 1906.12292}}].
	
	\bibitem{Johansson:2014zca}
	H.~Johansson and A.~Ochirov, \emph{{Pure Gravities via Color-Kinematics Duality
			for Fundamental Matter}},
	\href{https://doi.org/10.1007/JHEP11(2015)046}{\emph{JHEP} {\bfseries 11}
		(2015) 046} [\href{https://arxiv.org/abs/1407.4772}{{\ttfamily 1407.4772}}].
	
	\bibitem{Plefka:2019wyg}
	J.~Plefka, C.~Shi and T.~Wang, \emph{{Double copy of massive scalar QCD}},
	\href{https://doi.org/10.1103/PhysRevD.101.066004}{\emph{Phys. Rev. D}
		{\bfseries 101} (2020) 066004}
	[\href{https://arxiv.org/abs/1911.06785}{{\ttfamily 1911.06785}}].
	
	\bibitem{Bautista:2019tdr}
	Y.F.~Bautista and A.~Guevara, \emph{{From Scattering Amplitudes to Classical
			Physics: Universality, Double Copy and Soft Theorems}},
	\href{https://arxiv.org/abs/1903.12419}{{\ttfamily 1903.12419}}.
	
	\bibitem{Bautista:2019evw}
	Y.F.~Bautista and A.~Guevara, \emph{{On the double copy for spinning matter}},
	\href{https://doi.org/10.1007/JHEP11(2021)184}{\emph{JHEP} {\bfseries 11}
		(2021) 184} [\href{https://arxiv.org/abs/1908.11349}{{\ttfamily
			1908.11349}}].
	
	\bibitem{PV:2019uuv}
	A.~P.~V. and A.~Manu, \emph{{Classical double copy from Color Kinematics
			duality: A proof in the soft limit}},
	\href{https://doi.org/10.1103/PhysRevD.101.046014}{\emph{Phys. Rev. D}
		{\bfseries 101} (2020) 046014}
	[\href{https://arxiv.org/abs/1907.10021}{{\ttfamily 1907.10021}}].
	
	\bibitem{Cardoso:2016ngt}
	G.L.~Cardoso, S.~Nagy and S.~Nampuri, \emph{{A double copy for $ \mathcal{N}=2
			$ supergravity: a linearised tale told on-shell}},
	\href{https://doi.org/10.1007/JHEP10(2016)127}{\emph{JHEP} {\bfseries 10}
		(2016) 127} [\href{https://arxiv.org/abs/1609.05022}{{\ttfamily
			1609.05022}}].
	
	\bibitem{Cardoso:2016amd}
	G.~Cardoso, S.~Nagy and S.~Nampuri, \emph{{Multi-centered $ \mathcal{N}=2 $ BPS
			black holes: a double copy description}},
	\href{https://doi.org/10.1007/JHEP04(2017)037}{\emph{JHEP} {\bfseries 04}
		(2017) 037} [\href{https://arxiv.org/abs/1611.04409}{{\ttfamily
			1611.04409}}].
	
	\bibitem{Borsten:2020xbt}
	L.~Borsten and S.~Nagy, \emph{{The pure BRST Einstein-Hilbert Lagrangian from
			the double-copy to cubic order}},
	\href{https://doi.org/10.1007/JHEP07(2020)093}{\emph{JHEP} {\bfseries 07}
		(2020) 093} [\href{https://arxiv.org/abs/2004.14945}{{\ttfamily
			2004.14945}}].
	
	\bibitem{Anastasiou:2014qba}
	A.~Anastasiou, L.~Borsten, M.J.~Duff, L.J.~Hughes and S.~Nagy,
	\emph{{Yang-Mills origin of gravitational symmetries}},
	\href{https://doi.org/10.1103/PhysRevLett.113.231606}{\emph{Phys. Rev. Lett.}
		{\bfseries 113} (2014) 231606}
	[\href{https://arxiv.org/abs/1408.4434}{{\ttfamily 1408.4434}}].
	
	\bibitem{Anastasiou:2018rdx}
	A.~Anastasiou, L.~Borsten, M.~Duff, S.~Nagy and M.~Zoccali, \emph{{Gravity as
			Gauge Theory Squared: A Ghost Story}},
	\href{https://doi.org/10.1103/PhysRevLett.121.211601}{\emph{Phys. Rev. Lett.}
		{\bfseries 121} (2018) 211601}
	[\href{https://arxiv.org/abs/1807.02486}{{\ttfamily 1807.02486}}].
	
	\bibitem{Luna:2018dpt}
	A.~Luna, R.~Monteiro, I.~Nicholson and D.~O'Connell, \emph{{Type D Spacetimes
			and the Weyl Double Copy}},
	\href{https://doi.org/10.1088/1361-6382/ab03e6}{\emph{Class. Quant. Grav.}
		{\bfseries 36} (2019) 065003}
	[\href{https://arxiv.org/abs/1810.08183}{{\ttfamily 1810.08183}}].
	
	\bibitem{Alawadhi:2020jrv}
	R.~Alawadhi, D.S.~Berman and B.~Spence, \emph{{Weyl doubling}},
	\href{https://doi.org/10.1007/JHEP09(2020)127}{\emph{JHEP} {\bfseries 09}
		(2020) 127} [\href{https://arxiv.org/abs/2007.03264}{{\ttfamily
			2007.03264}}].
	
	\bibitem{Monteiro:2011pc}
	R.~Monteiro and D.~O'Connell, \emph{{The Kinematic Algebra From the Self-Dual
			Sector}}, \href{https://doi.org/10.1007/JHEP07(2011)007}{\emph{JHEP}
		{\bfseries 07} (2011) 007} [\href{https://arxiv.org/abs/1105.2565}{{\ttfamily
			1105.2565}}].
	
	\bibitem{Chacon:2020fmr}
	E.~Chac\'on, H.~Garc\'\i{}a-Compe\'an, A.~Luna, R.~Monteiro and C.D.~White,
	\emph{{New heavenly double copies}},
	\href{https://doi.org/10.1007/JHEP03(2021)247}{\emph{JHEP} {\bfseries 03}
		(2021) 247} [\href{https://arxiv.org/abs/2008.09603}{{\ttfamily
			2008.09603}}].
	
	\bibitem{Mizera:2018jbh}
	S.~Mizera and B.~Skrzypek, \emph{{Perturbiner Methods for Effective Field
			Theories and the Double Copy}},
	\href{https://doi.org/10.1007/JHEP10(2018)018}{\emph{JHEP} {\bfseries 10}
		(2018) 018} [\href{https://arxiv.org/abs/1809.02096}{{\ttfamily
			1809.02096}}].
	
	\bibitem{Cheung:2018wkq}
	C.~Cheung, I.Z.~Rothstein and M.P.~Solon, \emph{{From Scattering Amplitudes to
			Classical Potentials in the Post-Minkowskian Expansion}},
	\href{https://doi.org/10.1103/PhysRevLett.121.251101}{\emph{Phys. Rev. Lett.}
		{\bfseries 121} (2018) 251101}
	[\href{https://arxiv.org/abs/1808.02489}{{\ttfamily 1808.02489}}].
	
	\bibitem{Bern:2019nnu}
	Z.~Bern, C.~Cheung, R.~Roiban, C.-H.~Shen, M.P.~Solon and M.~Zeng,
	\emph{{Scattering Amplitudes and the Conservative Hamiltonian for Binary
			Systems at Third Post-Minkowskian Order}},
	\href{https://doi.org/10.1103/PhysRevLett.122.201603}{\emph{Phys. Rev. Lett.}
		{\bfseries 122} (2019) 201603}
	[\href{https://arxiv.org/abs/1901.04424}{{\ttfamily 1901.04424}}].
	
	\bibitem{Bern:2019crd}
	Z.~Bern, C.~Cheung, R.~Roiban, C.-H.~Shen, M.P.~Solon and M.~Zeng, \emph{{Black
			Hole Binary Dynamics from the Double Copy and Effective Theory}},
	\href{https://doi.org/10.1007/JHEP10(2019)206}{\emph{JHEP} {\bfseries 10}
		(2019) 206} [\href{https://arxiv.org/abs/1908.01493}{{\ttfamily
			1908.01493}}].
	
	\bibitem{Almeida:2020mrg}
	G.L.~Almeida, S.~Foffa and R.~Sturani, \emph{{Classical Gravitational
			Self-Energy from Double Copy}},
	\href{https://doi.org/10.1007/JHEP11(2020)165}{\emph{JHEP} {\bfseries 11}
		(2020) 165} [\href{https://arxiv.org/abs/2008.06195}{{\ttfamily
			2008.06195}}].
	
	\bibitem{Kalin:2019rwq}
	G.~K\"alin and R.A.~Porto, \emph{{From Boundary Data to Bound States}},
	\href{https://doi.org/10.1007/JHEP01(2020)072}{\emph{JHEP} {\bfseries 01}
		(2020) 072} [\href{https://arxiv.org/abs/1910.03008}{{\ttfamily
			1910.03008}}].
	
	\bibitem{Kalin:2020mvi}
	G.~K\"alin and R.A.~Porto, \emph{{Post-Minkowskian Effective Field Theory for
			Conservative Binary Dynamics}},
	\href{https://doi.org/10.1007/JHEP11(2020)106}{\emph{JHEP} {\bfseries 11}
		(2020) 106} [\href{https://arxiv.org/abs/2006.01184}{{\ttfamily
			2006.01184}}].
	
	\bibitem{Kalin:2020fhe}
	G.~K\"alin, Z.~Liu and R.A.~Porto, \emph{{Conservative Dynamics of Binary
			Systems to Third Post-Minkowskian Order from the Effective Field Theory
			Approach}}, \href{https://doi.org/10.1103/PhysRevLett.125.261103}{\emph{Phys.
			Rev. Lett.} {\bfseries 125} (2020) 261103}
	[\href{https://arxiv.org/abs/2007.04977}{{\ttfamily 2007.04977}}].
	
	\bibitem{Huang:2019cja}
	Y.-T.~Huang, U.~Kol and D.~O'Connell, \emph{{Double copy of electric-magnetic
			duality}}, \href{https://doi.org/10.1103/PhysRevD.102.046005}{\emph{Phys.
			Rev. D} {\bfseries 102} (2020) 046005}
	[\href{https://arxiv.org/abs/1911.06318}{{\ttfamily 1911.06318}}].
	
	\bibitem{Banerjee:2019saj}
	A.~Banerjee, E.O.~Colg\'ain, J.A.~Rosabal and H.~Yavartanoo, \emph{{Ehlers as
			EM duality in the double copy}},
	\href{https://doi.org/10.1103/PhysRevD.102.126017}{\emph{Phys. Rev. D}
		{\bfseries 102} (2020) 126017}
	[\href{https://arxiv.org/abs/1912.02597}{{\ttfamily 1912.02597}}].
	
	\bibitem{Casali:2020vuy}
	E.~Casali and A.~Puhm, \emph{{Double Copy for Celestial Amplitudes}},
	\href{https://doi.org/10.1103/PhysRevLett.126.101602}{\emph{Phys. Rev. Lett.}
		{\bfseries 126} (2021) 101602}
	[\href{https://arxiv.org/abs/2007.15027}{{\ttfamily 2007.15027}}].
	
	\bibitem{Adamo:2017nia}
	T.~Adamo, E.~Casali, L.~Mason and S.~Nekovar, \emph{{Scattering on plane waves
			and the double copy}},
	\href{https://doi.org/10.1088/1361-6382/aa9961}{\emph{Class. Quant. Grav.}
		{\bfseries 35} (2018) 015004}
	[\href{https://arxiv.org/abs/1706.08925}{{\ttfamily 1706.08925}}].
	
	\bibitem{Ilderton:2018lsf}
	A.~Ilderton, \emph{{Screw-symmetric gravitational waves: a double copy of the
			vortex}}, \href{https://doi.org/10.1016/j.physletb.2018.04.069}{\emph{Phys.
			Lett. B} {\bfseries 782} (2018) 22}
	[\href{https://arxiv.org/abs/1804.07290}{{\ttfamily 1804.07290}}].
	
	\bibitem{Adamo:2018mpq}
	T.~Adamo, E.~Casali, L.~Mason and S.~Nekovar, \emph{{Plane wave backgrounds and
			colour-kinematics duality}},
	\href{https://doi.org/10.1007/JHEP02(2019)198}{\emph{JHEP} {\bfseries 02}
		(2019) 198} [\href{https://arxiv.org/abs/1810.05115}{{\ttfamily
			1810.05115}}].
	
	\bibitem{Adamo:2019zmk}
	T.~Adamo and A.~Ilderton, \emph{{Gluon helicity flip in a plane wave
			background}}, \href{https://doi.org/10.1007/JHEP06(2019)015}{\emph{JHEP}
		{\bfseries 06} (2019) 015}
	[\href{https://arxiv.org/abs/1903.01491}{{\ttfamily 1903.01491}}].
	
	\bibitem{Adamo:2020qru}
	T.~Adamo and A.~Ilderton, \emph{{Classical and quantum double copy of
			back-reaction}}, \href{https://doi.org/10.1007/JHEP09(2020)200}{\emph{JHEP}
		{\bfseries 09} (2020) 200}
	[\href{https://arxiv.org/abs/2005.05807}{{\ttfamily 2005.05807}}].
	
	\bibitem{White:2016jzc}
	C.D.~White, \emph{{Exact solutions for the biadjoint scalar field}},
	\href{https://doi.org/10.1016/j.physletb.2016.10.052}{\emph{Phys. Lett.}
		{\bfseries B763} (2016) 365}
	[\href{https://arxiv.org/abs/1606.04724}{{\ttfamily 1606.04724}}].
	
	\bibitem{DeSmet:2017rve}
	P.-J.~De~Smet and C.D.~White, \emph{{Extended solutions for the biadjoint
			scalar field}},
	\href{https://doi.org/10.1016/j.physletb.2017.11.007}{\emph{Phys. Lett.}
		{\bfseries B775} (2017) 163}
	[\href{https://arxiv.org/abs/1708.01103}{{\ttfamily 1708.01103}}].
	
	\bibitem{Bahjat-Abbas:2018vgo}
	N.~Bahjat-Abbas, R.~Stark-Much\~ao and C.D.~White, \emph{{Biadjoint wires}},
	\href{https://doi.org/10.1016/j.physletb.2018.11.026}{\emph{Phys. Lett. B}
		{\bfseries 788} (2019) 274}
	[\href{https://arxiv.org/abs/1810.08118}{{\ttfamily 1810.08118}}].
	
	\bibitem{Kodama:2000fa}
	H.~Kodama, A.~Ishibashi and O.~Seto, \emph{{Brane world cosmology: Gauge
			invariant formalism for perturbation}},
	\href{https://doi.org/10.1103/PhysRevD.62.064022}{\emph{Phys. Rev.}
		{\bfseries D62} (2000) 064022}
	[\href{https://arxiv.org/abs/hep-th/0004160}{{\ttfamily hep-th/0004160}}].
	
	\bibitem{Kodama:2003jz}
	H.~Kodama and A.~Ishibashi, \emph{{A Master equation for gravitational
			perturbations of maximally symmetric black holes in higher dimensions}},
	\href{https://doi.org/10.1143/PTP.110.701}{\emph{Prog. Theor. Phys.}
		{\bfseries 110} (2003) 701}
	[\href{https://arxiv.org/abs/hep-th/0305147}{{\ttfamily hep-th/0305147}}].
	
	\bibitem{Ishibashi:2004wx}
	A.~Ishibashi and R.M.~Wald, \emph{{Dynamics in nonglobally hyperbolic static
			space-times. 3. Anti-de Sitter space-time}},
	\href{https://doi.org/10.1088/0264-9381/21/12/012}{\emph{Class. Quant. Grav.}
		{\bfseries 21} (2004) 2981}
	[\href{https://arxiv.org/abs/hep-th/0402184}{{\ttfamily hep-th/0402184}}].
	
	\bibitem{Farrow:2018yni}
	J.A.~Farrow, A.E.~Lipstein and P.~McFadden, \emph{{Double copy structure of CFT
			correlators}}, \href{https://doi.org/10.1007/JHEP02(2019)130}{\emph{JHEP}
		{\bfseries 02} (2019) 130}
	[\href{https://arxiv.org/abs/1812.11129}{{\ttfamily 1812.11129}}].
	
	\bibitem{Fitzpatrick:2014vua}
	A.~Fitzpatrick, J.~Kaplan and M.T.~Walters, \emph{{Universality of
			Long-Distance AdS Physics from the CFT Bootstrap}},
	\href{https://doi.org/10.1007/JHEP08(2014)145}{\emph{JHEP} {\bfseries 08}
		(2014) 145} [\href{https://arxiv.org/abs/1403.6829}{{\ttfamily 1403.6829}}].
	
	\bibitem{Kaviraj:2015xsa}
	A.~Kaviraj, K.~Sen and A.~Sinha, \emph{{Universal anomalous dimensions at large
			spin and large twist}},
	\href{https://doi.org/10.1007/JHEP07(2015)026}{\emph{JHEP} {\bfseries 07}
		(2015) 026} [\href{https://arxiv.org/abs/1504.00772}{{\ttfamily
			1504.00772}}].
	
	\bibitem{Kraus:2020nga}
	P.~Kraus, S.~Megas and A.~Sivaramakrishnan, \emph{{Anomalous dimensions from
			thermal AdS partition functions}},
	\href{https://doi.org/10.1007/JHEP10(2020)149}{\emph{JHEP} {\bfseries 10}
		(2020) 149} [\href{https://arxiv.org/abs/2004.08635}{{\ttfamily
			2004.08635}}].
	
	\bibitem{Ghosh:2020lel}
	J.K.~Ghosh, R.~Loganayagam, S.G.~Prabhu, M.~Rangamani, A.~Sivakumar and
	V.~Vishal, \emph{{Effective field theory of stochastic diffusion from
			gravity}}, \href{https://doi.org/10.1007/JHEP05(2021)130}{\emph{JHEP}
		{\bfseries 05} (2021) 130}
	[\href{https://arxiv.org/abs/2012.03999}{{\ttfamily 2012.03999}}].
	
	\bibitem{Higuchi:1986wu}
	A.~Higuchi, \emph{{Symmetric Tensor Spherical Harmonics on the $N$ Sphere and
			Their Application to the De Sitter Group SO($N$,1)}},
	\href{https://doi.org/10.1063/1.527513}{\emph{J. Math. Phys.} {\bfseries 28}
		(1987) 1553}.
	
\end{thebibliography}
\end{document}